\documentclass[sigconf]{acmart}
\usepackage{multirow}
\usepackage{multicol}
\usepackage{colortbl}
\usepackage{xcolor}
\usepackage{soul}
\usepackage{enumitem}
\usepackage{arydshln}
\usepackage{tikz}

\definecolor{lightblue}{HTML}{b0dcf4} 
\definecolor{lightgreen}{HTML}{d8ecc4} 
\definecolor{lightyellow}{HTML}{f8f4b0} 
\definecolor{DR_orange}{HTML}{ffe3c8} 
\definecolor{DR_blue}{HTML}{c2cae4} 
\definecolor{DR_pink}{HTML}{fecbe5} 
\definecolor{DR_bluegreen}{HTML}{c9ebe7} 
\definecolor{DR_yellowgreen}{HTML}{f6f6cf} 
\newcommand{\highlight}[2]{\colorbox{#1}{#2}}

\definecolor{mediumgray}{HTML}{808080}

\definecolor{VA}{HTML}{FFDFB5} 
\definecolor{design}{HTML}{F7F8A7}
\sethlcolor{design} 
\newcolumntype{M}{>{\ttfamily\small}p{14cm}}
\newcolumntype{L}{>{\raggedright\arraybackslash}p{2.7cm}}

\definecolor{custompurple}{HTML}{a275ab}
\definecolor{custompink}{HTML}{f27d9b}

\newcommand{\circlewithnumber}[4][1.5]{%
\tikz[baseline=(char.base)]{
    \node[shape=circle,fill=#2,draw=#2,inner sep=#1pt/3, text=white, opacity=#4] (char) {\small #3};}%
}

\newcolumntype{S}{>{\ttfamily\small}p{8.5cm}}
\newcolumntype{X}{>{\columncolor{VA}\ttfamily\small}p{8.5cm}}
\newcommand{\pquotes}[1]{\textcolor[gray]{0.35}{\textit{#1}}}
\newcommand{\tquotes}[1]{\textcolor[gray]{0.35}{\textit{#1}}}
\newcommand{\vquotes}[1]{{\fontfamily{lmr}\selectfont\small\textcolor[gray]{0.2}{\textit{#1}}}}

\def\eg{\emph{e.g., }} 
\def\ie{\emph{i.e., }}

\newcommand{\amh}[1]{ \textcolor{black}{#1}}
\newcommand{\vah}[1]{ \textcolor{black}{#1}}
\newcommand{\mws}[1]{ \textcolor{black}{#1}}
\newcommand{\sally}[1]{ \textcolor{black}{#1}}

\AtBeginDocument{%
  \providecommand\BibTeX{{%
    \normalfont B\kern-0.5em{\scshape i\kern-0.25em b}\kern-0.8em\TeX}}}

\copyrightyear{2025}
\acmYear{2025}
\setcopyright{cc}
\setcctype{by-nd}
\acmConference[CHI '25]{CHI Conference on Human Factors in Computing Systems}{April 26-May 1, 2025}{Yokohama, Japan}
\acmBooktitle{CHI Conference on Human Factors in Computing Systems (CHI '25), April 26-May 1, 2025, Yokohama, Japan}\acmDOI{10.1145/3706598.3713839}
\acmISBN{979-8-4007-1394-1/25/04}

\begin{document}


\title[Voice Assistants for Health Self-Management: Designing for and with Older Adults]{Voice Assistants for Health Self-Management: \\Designing for and with Older Adults}

\author{Amama Mahmood}
\email{amama.mahmood@jhu.edu}
\affiliation{%
  \institution{Johns Hopkins University}
  \streetaddress{3400 N. Charles St}
  \city{Baltimore}
  \state{Maryland}
  \country{USA}
  \postcode{21218}
}

\author{Shiye Cao}
\email{scao14@jhu.edu}
\affiliation{%
  \institution{Johns Hopkins University}
  \streetaddress{3400 N. Charles St}
  \city{Baltimore}
  \state{Maryland}
  \country{USA}
  \postcode{21218}
}

\author{Maia Stiber}
\email{mstiber@jhu.edu}
\affiliation{%
  \institution{Johns Hopkins University}
  \streetaddress{3400 N. Charles St}
  \city{Baltimore}
  \state{Maryland}
  \country{USA}
  \postcode{21218}
}

\author{Victor Nikhil Antony}
\email{vantony1@jhu.edu}
\affiliation{%
  \institution{Johns Hopkins University}
  \streetaddress{3400 N. Charles St}
  \city{Baltimore}
  \state{Maryland}
  \country{USA}
  \postcode{21218}
}

\author{Chien-Ming Huang}
\email{chienming.huang@jhu.edu}
\affiliation{%
  \institution{Johns Hopkins University}
  \streetaddress{3400 N. Charles St}
  \city{Baltimore}
  \state{Maryland}
  \country{USA}
  \postcode{21218}
}

\renewcommand{\shortauthors}{Mahmood, et al.}

\begin{abstract}
Supporting older adults in health self-management is crucial for promoting independent aging, particularly given the growing strain on healthcare systems. While voice assistants (VAs) hold the potential to support aging in place, they often lack tailored assistance and present usability challenges. We addressed these issues through a five-stage design process with older adults to develop a personal health assistant. 
Starting with in-home interviews ($N=17$), we identified two primary challenges in older adult's health self-management: health \textit{awareness} and medical \textit{adherence}. To address these challenges, we developed a high-fidelity LLM-powered VA prototype to debrief doctor’s after-visit summary and generate tailored medication reminders. We refined our prototype with feedback from co-design workshops ($N=10$) and validated its usability through in-home studies ($N=5$). Our work highlights key design features for personal health assistants and provides broader insights into desirable VA characteristics, including personalization, adapting to user context, and respect for user autonomy.

\end{abstract}





\begin{teaserfigure}
  \includegraphics[width=\textwidth]{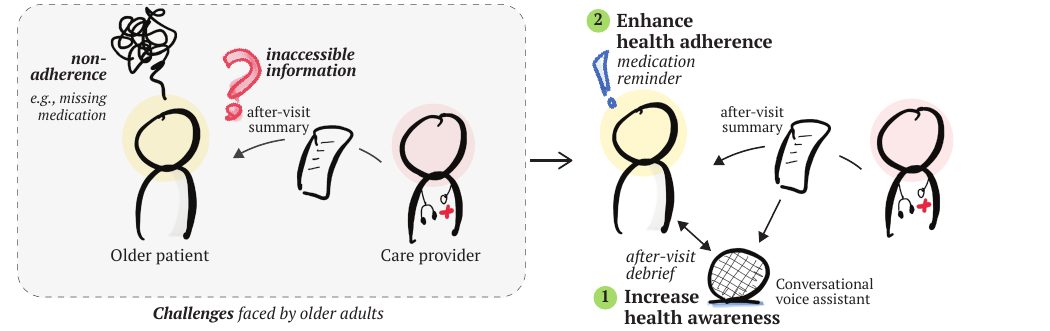}
  \caption{In this work, we identify and address two main challenges in older adults' health self-management: 1) health \textbf{awareness} and 2) \textbf{adherence} to medical regimens. We co-designed an LLM-powered voice assistant with older adults that uses doctors' after-visit summary (AVS) to provide debriefs and facilitates the creation of medication reminders.}
  \Description{The figure uses icons or simple shapes to represent key components, such as the older patient, the care provider, after-visit summary and the conversational assistant. The relationships between these elements are shown through arrows or connecting lines that highlight the flow of information and interactions. This figure has two blocks. The block on left titled ``challenges faced by older adults'' that shows a care provider gives after-visit summary to the older patient; this link is labeled with ``inaccessible information'' as first challenges. The older patient shows confusion such as missing medications which is labelled with ``non-adherence'' as second challenge.
  The second block shows the same link of doctor passing the after-visit summary to the patient. Now the after-visit summary is also passed to the conversational voice assistant and the conversational voice assistant has a bi-direction communication with the older patient which is labeled with first solution ``after-visit debrief''. This link is also labeled with first benefit for older adults ``increase health awareness''. The older patient is now also labeled with second solution ``medication reminder''. That is also labeled to show the second benefit for older patient ``enhance health adherence.''}
  \label{fig:teaser}
\end{teaserfigure}
\maketitle

\section{Introduction}

Aging gracefully is everyone's desire yet chronic health conditions often stand in the way. 
Nearly 95\% of the older adult population suffers from at least one chronic ailment, 
and 80\% endure two or more \cite{ncoa2022chronic}. 
Managing multiple chronic conditions is a formidable task for the aging population \cite{schoenberg2009s, upshur2008chronicity}, especially for those who live in social isolation and lack a support system---a reality for 31\% of women and 19\% of men aged 65 and older \cite{aging2020}.
Cognitive impairment, affecting one in five older adults \cite{pais2020global}, further hinders their ability to independently follow health management routines by impairing their understanding, memory, decision-making, and the ability to follow instructions.
These challenges are made even more pressing by rapidly growing aging population---by 2030, one in six people globally will be over 60, and the number is expected to double by 2050 \cite{aging2020WHO}. This demographic shift is straining caregivers and healthcare systems, which are increasingly overwhelmed by the demands of caring for older adults \cite{bardach2011role, giovannetti2012difficulty, shahly2013cross, mitchell2020global}.  

In this landscape, promoting self-care and self-management is crucial for older adults' well-being and for supporting aging in place \cite{davies2011promoting, mcleroy1988ecological, garnett2018self}. 
The importance of self-management is underscored by WHO's introduction of mAgeing (mobile aging), which aims to help older adults maintain functional ability and live as independently and healthily as possible through evidence-based self-management and self-care interventions at a national level \cite{who2018mageing}.
These interventions can empower individuals to take charge of their health, promoting independence and better overall health \cite{helbostad2017mobile}. 
\amh{Mobile health (mHealth) applications have the potential to support self-management but have poor usability, are overly complex, and fail to provide effective personalization, making them unsuitable for older adults \cite{wildenbos2019mobile, patel2020prospective, pater2017addressing}.}

Voice assistants (VAs) are a promising alternative to support autonomy and improve quality of life \cite{smith2023smart, blair2019understanding} as older adults prefer speech-based over text-based interfaces \cite{kowalski2019older, wulf2014hands}. \mws{Prior work has shown that VAs can be used for both medical information seeking (\eg \cite{brewer2022empirical}) and medical reminders (\eg \cite{shade2020voice}).
However, despite their potential to serve as health aids \cite{bolanos2020adapting, harrington2022s, sanders2019exploring, chen2021understanding}, VAs introduce usability challenges, as they are not designed with older adults in mind \cite{stigall2019older}. For instance, older adults find it difficult to create and manage event-specific reminders, such as appointment and medication reminders \cite{mahmood2024situated, shade2020voice, bolanos2020adapting}. }
\mws{Additionally, similar to mHealth applications, VAs are unable to personalize due to lack of knowledge in users' health context and preferences---both necessary to support health self-management \cite{takagi2018evaluating}---leaving the burden on users \cite{jesus2020voice}.}

To address these issues in VAs for health self-management, we conducted a five-stage design study with older adults \amh{to develop a personalized health assistant}. We started with in-home interviews ($N=17$) to learn more about older adults' perspectives on health management. \mws{From these interviews, we identified two primary challenges: older adults have difficulty 1) staying aware of their health status and 2) adhering to health regimens.} Based on these insights, we designed and developed a high-fidelity prototype VA by integrating a Large Language Model (LLM)---GPT-4---into a commercial VA (Alexa). The prototype focuses on three key phases: 1) processing patient's \amh{after-visit summary (AVS)} to understand their medical context, 2) debriefing user on the information from the AVS to enhance awareness, and 3) creating tailored medication reminders to improve adherence. To further refine the prototype, we held three co-design workshops ($N=10$) where participants provided actionable considerations based on their interaction with the initial prototype\mws{, such as clustering medications by time and adapting to regular and irregular routines.} We used the insights gathered to refine the prototype. Finally, we evaluated the usability of this refined VA using older adults'  health data ($N=5$) and found a high mean usability score of 85 as measured by the System Usability Scale (SUS) \cite{brooke1996sus}, with all participants successfully navigating the after-visit debrief (phase 2) and medication reminder creation (phase 3). 
Our work has the following contributions:

\begin{itemize}  [leftmargin=*, nolistsep]

    \item \textbf{Challenges} and \textbf{breakdowns} that the aging population faces in self-management of chronic diseases (in-home interviews; Section \ref{sec:interviews}).

    \item \textbf{LLM-powered personal health assistant} to address \textit{awareness} and \textit{adherence} challenges to support older adults in self-management through personalization (multi-stage design process; Section \ref{sec:protoype-VA}, \ref{sec:co-design-workshops}, and \ref{sec:refinement-VA}).

    \item \mws{\textbf{Empirical evidence}} demonstrating the \sally{usability} of our personal health assistant in debriefing the AVS, handling personal health-related queries and setting personalized reminders (\mws{in-home validation study;} Section \ref{sec:evaluation-study}).  
\end{itemize}

\section{Related Work}
We review technologies to support older adults' health self-management, with focus on role of voice assistants.
\subsection{Technologies for Older Adults' Health Self-Management}

\vah{Managing medication routines is a common challenge for older adults, compounded by age-related changes in vision, hearing, and dexterity. Several mHealth applications (\eg MyMedRec, DrugHub, PocketPharmacist) assist older adults by providing reminders, medication tracking, and facilitating communication with caregivers. While many apps feature basic medication reminders, they lack advanced functionalities such as automated medication tracking or detailed drug interaction information necessary for dynamic health routines \cite{grindrod2014evaluating}. Moreover, these tools often fail to overcome age-related challenges such as hearing or vision impairments and feature non-intuitive user interfaces. Setting up these reminders is also a significant barrier, as older adults often find the processes confusing, time-intensive, and incompatible with their established routines and existing medication management strategies \cite{pater2017addressing}.}

\vah{Wearable devices such as smartwatches, smart glasses, and smart pillboxes offer automated tracking and opportunities for situated intelligence \cite{deutsch2016smartwatch, xu2016medhelp, suzuki2014smartphone}. These technologies enable features such as real-time medication reminders or augmented guidance. However, physical interaction with such devices—such as manipulating small buttons or aligning sensors—can pose challenges for individuals with arthritis or reduced dexterity. Furthermore, the high cost, steep learning curve, and lack of contingency planning for events such as missed doses remain significant barriers to adoption \cite{mathur2022collaborative, pater2017addressing, patel2020prospective, martin2019exploring}. Importantly, existing solutions often fail to adapt to the entrenched routines of older adults. Many prefer to integrate new technologies with existing methods such as pillboxes, sticky notes, or paper calendars rather than replace them entirely, highlighting the importance of routine familiarity and ease of use \cite{mathur2022collaborative}.}

\vah{Voice-based conversational AI systems have the potential to overcome physical and visual barriers, and enable personalized, and context-aware interactions tailored to user needs and preferences \cite{chen2021understanding, even2022benefits, kocaballi2022design}. These capabilities position VAs as a compelling solution to advancing health management technologies for older adults. }

\subsection{Current Voice Assistants to Support Older Adults' Health}

Voice user interfaces (VUIs) show promise in supporting older adults by offering a natural, hands-free way to interact with digital information \cite{kumah2018electronic}. Speech input is three times faster and has a 20\% lower error rate than typing \cite{ruan2016speech}, which is especially beneficial for older adults as they may have physical or technological limitations. Hence, older adults prefer speech-based over text-based interfaces \cite{kowalski2019older, wulf2014hands}. VAs have shown significant potential in various aspects of health self-management, improving older adults' quality of life and autonomy \cite{smith2023smart, blair2019understanding}. They assist in medical information seeking \cite{brewer2022empirical, harrington2022s, sanders2019exploring} and setting reminders for management of pain and other medical conditions \cite{shade2020voice, bolanos2020adapting, chen2021understanding,  jesus2020voice}. Despite their potential in healthcare applications, commercial VAs are not specifically designed for older adults \cite{stigall2019older} and two major challenges that hinder their effectiveness: 1) usability and 2) lack of user context.

\subsubsection{Usability challenges}
\label{sec:background-usability}
Voice assistants are marketed and designed to mimic human-like conversations \cite{harris2004voice, gilmartin2017social}. Therefore, users may treat VAs as if they are engaging in human dialogue \cite{cowan2017can, doyle2019mapping}, specifically older adults \cite{mahmood2024situated}. Inaccurate user mental models of VA capabilities, driven by the notion of humanlike interactions, lead to overestimation of conversational abilities and communication breakdowns when expectations are unmet  \cite{luger2016like, leahu2013categories, moore2016progress}. Such misconceptions have been shown to impact older adults' health-related queries \cite{mahmood2024situated, brewer2022empirical}. For instance, a detailed analysis of health information-seeking behaviors shows that human-like expectations often lead to conversational breakdowns, such as VAs providing incorrect or off-topic answers \cite{brewer2022empirical}. These usability challenges are exacerbated for marginalized populations \eg people with cultural differences \cite{harrington2018designing} and physical disabilities \cite{pradhan2018accessibility}. 

Usability challenges are also prevalent in available functional features such as setting reminders.
Although commercial VAs offer reminder setting functionality, it is sub-optimal \cite{mahmood2024situated}. 
For instance, older adults often struggle to set recurring medication reminders, requiring them to set them each morning \cite{mahmood2024situated},  which defeats the purpose as they still have to rely on their memory every day.
This manual process \cite{mahmood2024situated, shade2020voice, bolanos2020adapting} leaves room for errors that can impact critical tasks such as attending doctor's appointments or taking medications.
Existing reminder features lack intuitiveness and personalization leaving them wanting a dedicated medication reminder function \cite{chen2021understanding}.
A VA that automatically creates reminders for older adults could be highly beneficial. For instance, the Snips VA provides reminders for medications and appointments but relies on manual data entry by healthcare providers \cite{jesus2020voice}. This process not only increases the workload for care providers but also makes older adults dependent for yet another task, which may be neither desirable nor practical. Thus, no current VA supports personalized reminder creation for medication and appointment self-management. 

\subsubsection{Lack of user context}
The lack of user context in interactions leads to more usability issues, particularly with current commercial VAs that treat each user query as a standalone event \cite{beirl2019using, kim2021exploring, liao2018all}. These VAs do not account for even basic conversation history, which could serve as a minimalistic proxy for user context. Implementing multi-turn conversations with large language models (LLMs) has been shown to enable more robust and fluid interactions with users \cite{mahmood2025user, chan2023mango, mahmood2024situated}, especially in health information-seeking scenarios, and helps reduce conversational breakdowns. However, this level of context may still fall short of fully understanding and supporting the needs and expectations of older adults from VA to support self-management. To the best of our knowledge, no VA currently incorporates personal health data for health queries or setting reminders. 
This work explores designing a VA that integrates user context---conversation history, daily routines, and personal health information---to better support older adults in self-management.

\subsection{LLM-Driven VAs for Health Self-Management.} \vah{LLMs have introduced advanced capabilities in context understanding and conversation generation \cite{gandhi2024understanding, peng2024customising}. When integrated into conversational agents, LLMs can facilitate more fluid and natural conversations and address usability and context-awareness challenges faced by current VAs \cite{mahmood2025user}. For instance, LLM-powered VAs enhance communication with older adults and improve health information collection while reducing provider time and effort \cite{yang2024talk2care}. LLMs have the potential to transform VAs into more effective tools for self-management, enabling seamless creation, execution, and tracking of context-aware reminders through a natural conversational interface. While prior work has explored LLM-driven chatbot architectures for chronic disease management \cite{montagna2023data} (\eg diabetes \cite{dao2024llm} and Parkinson’s \cite{cardenas2024autohealth}), to the best of our knowledge, no prior work has specifically explored how LLM-powered VAs can support self-management for older adults, specifically for medications.}

\begin{figure*}[th]
  \includegraphics[width=\textwidth]{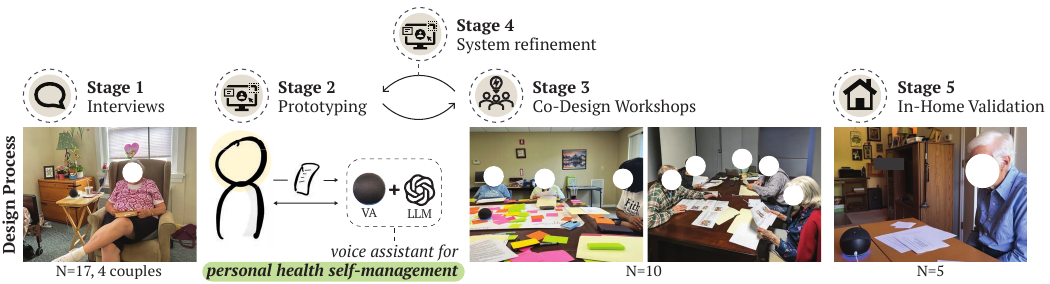}
  \caption{We designed personal health voice assistant to support older adults' self-management in 5-stages: stage 1---interviews with older adults; stage 2---initial prototyping; stage 3---co-design workshops with older adults; stage 4---refining the VA prototype; and stage 5---in-home validation study with older adults. All activities with participants were approved by our institutional review board.}
  \Description{The figure represents five stages of our design process from left to right. The first stage has a picture of older adult in their home with label stage 1 interviews. Underneath it the total number of participants is written as $N=17$, 4 couples. The Stage 2 is labeled as prototype and shows a representation of older adults interacting with the VA that is powered by an LLM and receives the AVS from the older adult. The VA and LLM together are labeled as voice assistant for personal health self-management. The stage three is labeled as co-design workshops and shows two photos from the co-design workshops with 3 and 4 participants in each sitting around a table with an experimenter. The pictures show the VA, personas, and collaborative maps on the tables. It is also labeled with $N=10$. The stage 2 and 3 are connected with circular errors one originating from stage 2 ending in stage 3 and one originating from stage 3 and ending in stage 2. The latter arrow is labeled with stage 4: system refinement. Finally on the right most, the stage 5 is labeled as in-home validation with a picture of a person talking to the VA and labeled $N=5$}
  \label{fig:process}
\end{figure*}

\subsection{Older Adults' Vision for Voice Assistants in Health Self-Management}

Older adults have expressed clear expectations for voice assistants in managing their health and well-being, particularly in the context of self-care and self-management. 
Older adults believe that VAs could offer a more comprehensive overview of their health, allowing them to access a ``holistic health'' summary that includes updates on their health progress \cite{martin2019exploring, pradhan2020use, harrington2018designing}. Their expectations for VAs extend beyond basic know-how about their health to include preventive care, with a desire for VAs to play a proactive role in monitoring health indicators and offering timely advice \cite{yamada2018development}. For instance, older adults have noted that they would appreciate being ``addressed by the smart speaker'' proactively and having the VA implement a ``watch-over'' function to ensure their well-being \cite{yamada2018development}.

Older adults emphasize the importance of VAs being easy to set up, intuitive to use, and robust enough to avoid frequent breakdowns \cite{chen2021understanding}. They expect VAs to handle periodic health management tasks with minimal user effort \cite{chen2021understanding}. 
Additionally, they value the ability to  easily modify reminder settings \cite{takagi2018evaluating}. 
They strongly desire control over their health data and decision-making, as well as transparency in data usage \cite{sanders2019exploring, chen2021understanding}.
These expectations of older adults, as articulated through interviews and observational studies, provide valuable insights into their envisioned VA for self-management. However, these insights often lack focus on practical aspects of design and implementation.
Therefore, in this work,  we involved older adults as co-designers to develop an LLM-powered VA that uses their personal health data and preferences to provide tailored support for after-visit debriefs and medication reminders.

\section*{Research Overview}
We involved older adults in a multi-stage design process (see Fig. \ref{fig:process}) to develop 
a voice assistant to support self-management. The following sections detail the design process, which includes five stages:
\textbf{1)} in-home interviews with older adults to identify breakdowns and challenges in their current health management practices (Section \ref{sec:interviews}), 
\textbf{2)} in-lab development of an initial VA prototype 
(Section \ref{sec:protoype-VA}), 
\textbf{3)} co-design workshops with older adults to gather design considerations for the VA (Section \ref{sec:co-design-workshops}), 
\textbf{4)} refining the initial prototype based on co-design findings (Section \ref{sec:refinement-VA}), and 
\textbf{5)} an in-home validation study to assess usability of the refined VA (Section \ref{sec:evaluation-study}). All activities with older adults were approved by our institutional review board and all participants were compensated at a rate of 15\$/hr.

\section{Stage 1: In-Home Interviews with Older Adults}
\label{sec:interviews}
We conducted semi-structured interviews\footnote{Stage 1 interview guide: \url{https://tinyurl.com/bdfw8n5c}. The URL contains all supplementary materials labeled under respective stages 1–5.} to explore the challenges the aging population encounters in managing their health, focusing on comprehending  medical information, remembering appointments, taking medications, and other critical aspects. 
The goal was to capture their personal narratives and perspectives on these challenges, excluding those of healthcare providers.

\subsection{Participants}
We interviewed 21 participants aged 66--94 ($M=74.71, SD = 3.83$)---9 females, 4 males, and 4 heterosexual couples. Among them, nine lived in a community center that has both an independent
($n=6$) and an assisted ($n=3$) living community, and 12 were community-dwelling adults. 
Table \ref{tab:participants} in appendix presents their demographics (P1--P17). 
We considered couples as single data point, since we interviewed them togetherleading to 17 data points comprising of 3 hours 50 minutes of audio data. 

\subsection{Data Analysis}
We transcribed all the interviews and conducted reflexive thematic analysis. Each transcript was analyzed independently by at least two members of the research team. 
The research team then convened to extensively discuss and reach consensus on the key breakdowns in participants' self-management
and the design requirements needed to support it.
Two key aspects of self-management, as detailed below, emerged: 1) awareness---understanding AVSs and test results and 2) adherence---managing appointments and medications.

\begin{table*}[tb]
\centering
\caption{Findings of interviews: breakdowns in older adults health self-management. Note: P stands for participant.}
\label{tab:findings-interview}
\begin{tabular}{c p{4.5cm} p{11.9cm}}

& \textbf{Breakdowns} & \textbf{Evidence from participant quotes} \\
\midrule
\midrule
\multirow{12}{*}{\rotatebox[origin=c]{90}{\textbf{Awareness}}}  & \multicolumn{2}{l}{\textbf{Comprehending after visit summary and test results}}  \\
& 1. Do not read or understand info & P8: \tquotes{``I don't read the follow up information and sometimes when I talk to the doctor or I have a visiting nurse, it's something I should have done.''} 
\vspace{1mm}
\\
& 2. Reliance on medical professionals & P15: \tquotes{``I also ask a lot of questions now because there was a time when the doctors 
\dots were not that great about things like that. And I ended up having some health issues because they didn't and I didn't really take care of me\dots''} 
\vspace{1mm}
\\
& 3. Incomplete information necessitates seeking additional info & P2: \tquotes{``I've been put on some medicine recently that \dots I recognize it's new medicine and so \dots I googled it.''}
\vspace{1mm}
\\
& 4. Deteriorating vision hinders interaction & P13: \tquotes{``It's becoming more and more laborious to go through the \dots visual material, whether it's in print or on the, on the tablet.''} 
\vspace{1mm}
\\
\midrule
\midrule

\multirow{20}{*}{\rotatebox[origin=c]{90}{\textbf{Adherence}}} & \multicolumn{2}{l}{\textbf{Managing doctor's appointments}}  \\
& 1. Forgetfulness &  P17: \tquotes{``I was supposed to have the stress test, and I was already to go, and for some reason, I lollygag around or whatever. And I didn't get out the door in time. I was 20 minutes late. I didn't call to say I'd be late, but they didn't take me. So I had to reschedule that.''} 
\vspace{1mm}
\\
& 2. Errors in creating calendar events or forgetting to set reminders & 
P9: \tquotes{``He [a friend] was headed for and he had borrowed his girlfriend's car to go to the appointment and he had a wreck. And so he called and said, I'm gonna be delayed\dots and they said that's ok. It isn't until tomorrow.''}
\vspace{1mm}
\\
& 3. Rescheduling is hard & P11b: \tquotes{``You have to wait a few months.''}
\vspace{1mm}
\\
\cline{2-3}
& \multicolumn{2}{l}{\textbf{Managing medications}}  \\
& 1. Trouble keeping track of intake & P10a: \tquotes{``It's hard to do it. It's hard to keep track of that specific [medication].''}  
\vspace{1mm}
\\
& 2. Forgetting doses outside of habitual routine & P14:  
\tquotes{``Sometimes, I'll put it in my phone to remind me to take my acid reflex pill, for instance, because I usually don't take that in the morning. I usually take that during the afternoon and then sometimes I forget and then I get acid reflex, then I'll say, oh, I didn't take my pill. ''} 
\vspace{1mm}
\\
& 3. Hard to know and keep track of medication interactions & P5: \tquotes{``I do run into trouble, like, when I have a complicated illness and I have to take multiple things that, because of counter interactions, I'm thinking of getting the nursing service here to come in and help me with that when it happens.''}. 
\vspace{1mm}
\\
& 4. Dependence on nursing staff &  P3: \tquotes{``Well sometimes they're not on time''}
\vspace{1mm}
\\
\hline
\end{tabular}
\end{table*}

\subsection{Findings: Challenges in Health Self-Management }

We identified two main challenges in participants' health self-management: 1) Awareness of health and 2) Adherence to medical regimens. The findings are summarized in Table \ref{tab:findings-interview}.

\begin{figure*}[th!]
  \includegraphics[width=\textwidth]{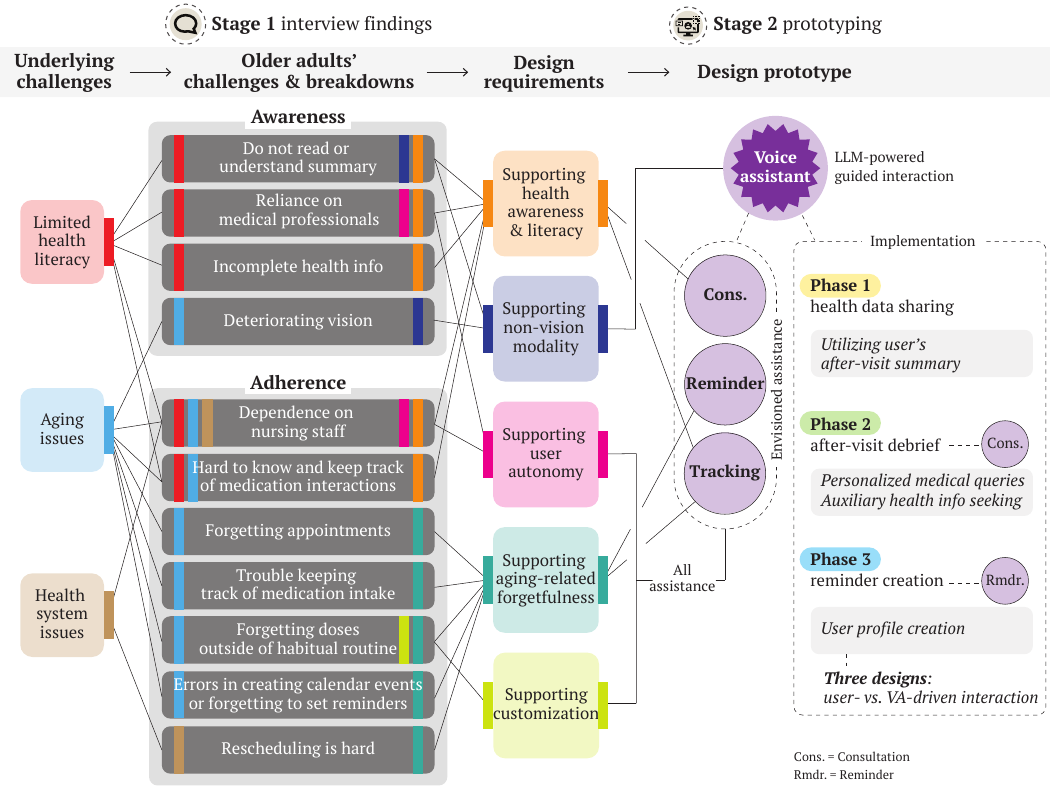}    
  \caption{We present the interview findings alongside their underlying causes. Based on these findings, we outline the design requirements and demonstrate their implementation in our initial prototype.} 
  \Description{The figure visualizes the findings of interviews and link them to the underlying challenges which then informs (links to) the design requirements that are then linked to the design prototype we develop. 
  First from underlying challenges to older adults' challenges and breakdowns are: 
  1. Limited Health Literacy to Do not read or understand summary, Reliance on medical professionals, Incomplete health information, Dependence on nursing staff, and Hard to know and keep track of medication interactions. 
  2. Aging issues to deteriorating vision, dependence on nursing staff, hard to know and keep track of medication interactions, forgetting appointments, Trouble keeping track of medication intake, Forgetting doses outside of habitual routine, and errors in creating calendar events or forgetting to set reminder. 
  3. Health system issues to Dependence on nursing staff and Rescheduling is hard. 
  Next is Older Adults’ Challenges and Breakdowns to Design Requirements: 
  1. Do Not Read or Understand Summary, Reliance on Medical Professionals, Incomplete Health Information, Dependence on Nursing Staff, and Hard to Know and Keep Track of Medication Interactions to Design requirement Supporting Health Awareness and Literacy
  2. Do Not Read or Understand Summary and Deteriorating vision to design requirement Supporting Non-Vision Modality
  3. Reliance on medical professionals and Dependence on Nursing Staff to design requirement Supporting User Autonomy. 
  4. Forgetting Appointments, Hard to Know and Keep Track of Medication Interactions, Forgetting Doses outside of habitual hours, Trouble Keeping Track of Medication intake, Errors in Creating Calendar Events or Forgetting to Set Reminders, and rescheduling is hard to design requirement Supporting Aging-Related Forgetfulness
  5. Forgetting Doses outside of habitual hours to design requirement Supporting Customization. 
  Next is Design requirements to design prototype: 
  1. Supporting non-vision modality to voice assistant (LLM-powered guided interaction)
  2.Supporting Health Awareness and Literacy to design prototype purpose consultation and tracking of envisioned assistance.
  3. Supporting aging-related forgetfulness to design prototype purpose reminder and tracking of envisioned assistance.
  4. Supporting user autonomy and customization is applied to over all assistance.
  The envisioned assistance has three purposes: consultation, reminder and tracking. The implementation block of the design prototype shows three phases: Phase 1 is health data sharing that utilizes user's AVS. Phase 2 is after-visit debrief that fulfills purpose of consultation with personalized medical queries and auxiliary health information seeking. Phase 3 is reminder creation that fulfills the purpose of reminder by creating user profile and have three design alternatives: user vs. VA driven interaction. 
  }
\label{fig:stage1-stage2}
\end{figure*}

\subsubsection{Awareness of health} 
Following are our findings on breakdowns in their awareness of health:\\
\textbf{Accessing and comprehending AVS and test results.}
In 2009, the United States began offering incentives to healthcare organizations to provide patients with clinical summaries after office visits, including personalized information on their diagnoses, medications, and upcoming appointments \cite{pathak2020patient}. Since then, while providing patients with an AVS is not a requirement, it has been widely adopted as a standard of care for outpatient visits \cite{hummel2012providing}. Our participants highlighted the importance of these AVSs when it comes to understanding their overall health, medications and lifestyle recommendations. 
However, we identified several challenges with how patient engagement with AVSs and test results affect their awareness.
First, participants may not read or fully comprehend important information within these summaries which can sometimes lead to adverse health outcomes. 
Second, they have to rely on doctors' explanations which, if inadequate, can cause health issues. 
Third, a gap in information provided by the healthcare professionals provoked some participants to consult medical resources online such as Google or WebMD to understand their medications better. 
Fourth, participants' deteriorating vision may hinder their ability to read AVSs and test results.

\subsubsection{Adherence to health regimen} Following are the sub-themes identified for adherence to healthcare regimen:

\noindent \textbf{Managing doctor's appointment.}
The interviews revealed a number of challenges faced by participants in managing their doctor's appointments, mainly revolving around the effective use of reminders and the efficient organization of appointment schedules. 
Participants missed appointments usually due to 1) forgetfulness and 2) errors in creating calendar events or forgetting to set reminders along with calendar event thus relying on manually checking the calendar everyday. 
Missing appointments often has high stakes, as it can take months to reschedule and adds significant costs to healthcare resources \cite{manfredi2017sms, oliver2019david, ellis2017demographic}. 
Different living conditions create logistical challenges, such as couples needing to avoid scheduling conflicts due to mutual reliance. \\
\textbf{Medication management.}
Participants identified four key challenges in adhering to their medication intake regimen. First, they had trouble keeping track of whether they took medications. 
Second, tracking becomes more challenging with medication doses that fall outside of habitual morning or night routines. 
Third, for the complicated interaction effects between multiple medications they may need additional assistance.
Lastly, participants in assisted living centers expressed concerns about inconsistent medication times administered by nursing staff, emphasizing the importance of maintaining control over their health management.

\subsection{From Challenges to Design Requirements for Personal Health Assistant}
Prior work identifies three key underlying challenges to effective self-management in older adults: 1) limited health literacy, 2) aging-related issues, and 3) health system shortcomings \cite{doyle2019managing, nguyen2022systematic}. Older adults with low health literacy are more prone to chronic diseases and adverse health behaviors \cite{cho2008effects}. Aging-related issues, such as cognitive and physical impairments, further hinder their ability to manage their health effectively \cite{cramm2013understanding, bayliss2007barriers}. Additionally, healthcare systems often fail to provide the necessary support for self-management in this population \cite{nguyen2022systematic, mitchell2020global}.
Our interviews reveal how these underlying challenges affect older adults' health self-management in practice (see Fig. \ref{fig:stage1-stage2}). For instance, limited health literacy manifests as difficulty in understanding their health information, while aging issues result in forgetting appointments and medications.
Based on these findings, we identified design requirements (DR) to inform the prototype of the envisioned personal health assistant (Fig. \ref{fig:stage1-stage2}):

\begin{itemize}[leftmargin=*, nolistsep] 
    \item \textbf{\highlight{DR_blue}{DR1}: Supporting non-vision modality.} Our findings show that an abundance of visual materials (\eg AVS, medication instructions, and test results) hinders older adults from engaging with their health information, and deteriorating vision further limits their ability to read. Thus, the assistant should support non-vision modality such as voice interactions. 
    \item \textbf{\highlight{DR_orange}{DR2}: Supporting health awareness and literacy through consultation.} The assistant should improve older adults' awareness of their health by making medical information accessible and understandable. Through consultation, it should enable personalized health-related queries. To achieve this, the assistant must access patient data.
    \item \textbf{\highlight{DR_bluegreen}{DR3}: Supporting aging-related forgetfulness through reminders and tracking.} The assistant should support aging-related forgetfulness through regular reminders for events such as  medications to enhance adherence. Additionally, it should enable users to track their health status over time. 
    \item  \textbf{\highlight{DR_pink}{DR4} Supporting user autonomy.} Participants expressed a desire for control and autonomy in managing their health; consistent with prior work \cite{sanders2019exploring, chen2021understanding}. Therefore, the assistant should not compromise user autonomy when providing support.
    \item \textbf{\highlight{DR_yellowgreen}{DR5}: Supporting customization.} Participants struggled with taking medications outside of their morning and night routines. With varying daytime routines, fixed assistance may not be effective. Thus, the assistant should support customization, for instance, tailoring reminders to user's schedule and medical needs.
    
\end{itemize}

\section{Stage 2: Initial Prototyping of Envisioned Personal Health Assistant}  
\label{sec:protoype-VA}

To support older adults' health self-management, we developed the \textit{envisioned} personal health assistant  based on the design recommendations (Fig. \ref{fig:stage1-stage2}) and prior work \cite{mahmood2025user, mahmood2024situated, haas2022keep,hwang2023rewriting}. 
As established by \highlight{DR_blue}{DR1} and older adults inherent preference for voice assistants over text-based systems \cite{wulf2014hands, kowalski2019older}, we chose to design a personal health VA. VAs have also shown potential as health aides, supporting aging in place \cite{shade2020voice, bolanos2020adapting, brewer2022empirical, harrington2022s, sanders2019exploring, chen2021understanding}, offering hands-free, intuitive interfaces ideal for those with physical or cognitive limitations \cite{wildenbos2019mobile}. 
Furthermore, our interview study revealed that older adults often make errors when creating calendar events and frequently forget to set reminders. Prior research also highlights these usability challenges with general-purpose VAs while creating medication reminders \cite{mahmood2024situated, shade2020voice, bolanos2020adapting}. To reduce errors in these sensitive tasks, guidance is crucial. By leveraging LLMs' contextual understanding \cite{mahmood2025user, chan2023mango}---proven effective in health contexts for older adults \cite{yang2024talk2care, jo2023understanding}---we implemented an LLM-powered VA to enable a guided interaction.

The end-to-end prototype is developed  by integrating GPT-4 into Amazon's Alexa via a custom Alexa skill hosted on an Echo Dot smart speaker to enable a guided interaction. The purpose of the initial prototype is to present it to the older adults in the next stage---the co-design workshop. 
Next, we describe three phases of the designed interaction (see Fig. \ref{fig:stage1-stage2}): 1) health data sharing, 2) after-visit debrief, and 3) medication reminder creation. To achieve the desired behavior, each phase was implemented within ChatGPT-Alexa skill using prompt engineering; the associated prompts are provided in supplementary materials\footnotemark[1].

\subsection{\highlight{lightyellow}{Phase 1} Health Data Sharing}

Phase 1 targets capturing patient health data. In the envisioned system, users would provide a photo of an AVS via web app.  
However, for the initial prototype, we use a mock AVS \sally{that is modeled off of a real AVS} to simulate the interaction.

\subsection{\highlight{lightgreen}{Phase 2} After-Visit Debrief (Consultation)}
\label{sec:prototype1-summary}
Phase 2 targets \highlight{DR_orange}{DR2}. 
The VA drives the debrief by summarizing key information from the AVS, including conditions discussed, medication changes, lifestyle recommendations, and upcoming appointments.  
As AVSs can be information-dense, it is important to avoid overwhelming users with verbose responses that disrupt conversational flow \cite{mahmood2025user}. To balance concise yet comprehensive responses  \cite{haas2022keep, mahmood2024situated, hwang2023rewriting}, 
we implement a hierarchical structure \cite{mahmood2025user} where the VA presents one aspect of the summary (\eg medical conditions) and allows the user to ask questions. The VA then provides relevant information before moving to the next aspect (\eg medication changes), proceeding step-by-step until the document is fully summarized. 
Users can interrupt the VA and ask questions at any time \highlight{DR_pink}{DR4}; with the VA primarily referencing the AVS and using the LLM to augment responses as needed, achieved through prompt engineering.

\subsection{\highlight{lightblue}{Phase 3} Medication Reminder Creation}
\label{sec:system-design-options}
Phase 3 targets \highlight{DR_bluegreen}{DR3}. Given the diverse needs of older adults for medication management (\eg varying numbers of medications and irregular routines) and their desire for control and autonomy in healthcare, we chose to bring various interaction designs to the co-design workshops with older adults to give them alternatives to think about. 
Therefore, we developed three different  interaction designs for medication reminder creation varying in degree of user autonomy \highlight{DR_pink}{DR4} and customization \highlight{DR_yellowgreen}{DR5}:

\subsubsection{Design 1: VA-driven default reminders}
This design takes a proactive approach, with the VA suggesting reminders based on the medication details in the AVS, without considering user routine. It assumes standard routines (\eg common meal times) to set reminder timings, then reviews them with the user. \amh{For medications that can be taken at any time, the VA randomly suggests a time.} Users can ask to adjust reminders. This design gives the VA most control and allows only for post-hoc personalization by the user.

\subsubsection{Design 2: User-driven customized reminders}
The VA prompts the user to create a profile that maps their routine and reminder preferences. It gathers detailed information about wake-up and bedtimes, meal times, and other periods when the user is typically home to take medication. The VA also inquires about preferred reminder times outside the routine  \amh{to accommodate for medications that do not have strict requirements,} and any other specific preferences for medication-related events.
This approach gives the user control over reminder timings, allowing the VA to suggest customized reminders based on the user profile.

\subsubsection{Design 3: Hybrid tailored reminders}
The third design adopts a hybrid approach, \sally{asking only about the parts of the user routines} relevant to their specific medications. For instance, if a medication must be taken with dinner, the VA asks about dinner time before suggesting a reminder (\textit{VA-driven}).  \amh{If there is no specified requirement, the VA asks user for preferred time to be consistent with Design 2.} This design focuses on efficiency, requesting only essential information to customize reminders and avoiding unnecessary questions. However, while it reduces upfront queries to build a user profile, it may miss broader preferences, as the decision to provide additional details is up to the user (\textit{user-driven}).

\section{Stage 3: Co-design Workshops with Older Adults}
\label{sec:co-design-workshops}
We conducted three co-design workshops with older adults to identify key design considerations for a personal health assistant. In the workshops, we presented an initial prototype to facilitate hands-on interactions, allowing participants to visualize potential solutions to their challenges and offer feedback on improving the design.

\subsection{Procedure} 
The co-design workshops followed this procedure:

\begin{enumerate} [leftmargin=*, nolistsep]
    \item \textit{Introduction.} The participants were introduced to the envisioned personal health VA that could debrief a summary of a doctor’s visit, answer related questions, and set medication reminders based on the AVS. 
    \item \textit{Personas.} The participants were shown five personas\footnotemark[1],  
    created from the demographics and characteristics of our interview participants, representing different health management practices, technological proficiency, and lifestyles. The purpose of introducing personas is \sally{to encourage participants to consider different perspectives, including those of others they know, alongside their own viewpoint, and to reduce barriers as participants can bring up their own experience through personas given the sensitive nature of task.} 
    \item \textit{Group discussion.} Participants then engaged in an open group discussion to outline their initial impressions of the envisioned VA, their preferences for its features, and potential barriers they might encounter. They also utilized collaborative map-making to develop concrete solutions for possible breakdowns. 
    \item \textit{Interaction with high-fidelity VA prototypes.} To further the discussion, participants then took turns interacting with the VA prototype for after-visit debrief (phase 2) 
    and medication reminder creation (phase 3; three alternatives)
    using \sally{a mock AVS to protect people's privacy}. Afterward, they provided detailed feedback on various aspects of the interaction and discussed design improvements.
\end{enumerate}

\subsection{Participants}

We recruited a total of 10 participants aged 65--88 ($M=77.60, SD = 6.45$) for 3 workshops---four community dwelling adults for the first workshop (2 male, 2 female), three independent living community center dwelling adults (3 female) for the second workshop, and three immigrants to the United States for the third workshop. \sally{Four out of ten participants were recruited from the pool of participants for Stage 1 (see details in Appendix Table \ref{tab:participants}).} Taking into account the comfort of the participants, the third workshop was run in two parts---one individual (female, W3-B) and one couple (female, W3-N and male, W3-Y).
We deliberately recruited participants with varying living conditions to capture diverse perspectives on potential challenges influenced by different aging environments. Immigrant participants were included to explore additional challenges, such as unfamiliarity with the healthcare system and language barriers, and how VAs might address them.
Detailed demographics, participants' use of VA technology and current health management practices are presented in  Tables \ref{tab:participants} and \ref{tab:tech-use} (Appendix). 
The workshops resulted in about 8 hours 40 minutes of audio data.

\subsection{Data Analysis}
We conducted thematic analysis on the transcribed audio data to identify existing and potential breakdowns in participants' current practices and capture solutions proposed by participants for these breakdowns. Additionally, \sally{we coded their interactions with the VA prototype for the types of queries users made and their preferences while creating medication reminders.}  
Second part of third workshop was conducted predominately in Mandarin since participants (W3-N and W3-Y) felt more comfortable speaking in Mandarin. The conversations were facilitated by a member of the research team and later translated into English. 
The first author coded all the workshop transcripts and classified the codes into themes and sub-themes. Two researchers who were present at the workshops 
fully agreed on the outcomes of coding.  
The themes are: 1) consultation, 2) medication reminders, 3) health tracking, and 4) general considerations for VA design. For each sub-theme, we present challenge-solution pairs and user preferences for the VA.

\begin{figure*}[t]
     \includegraphics[width=\textwidth]{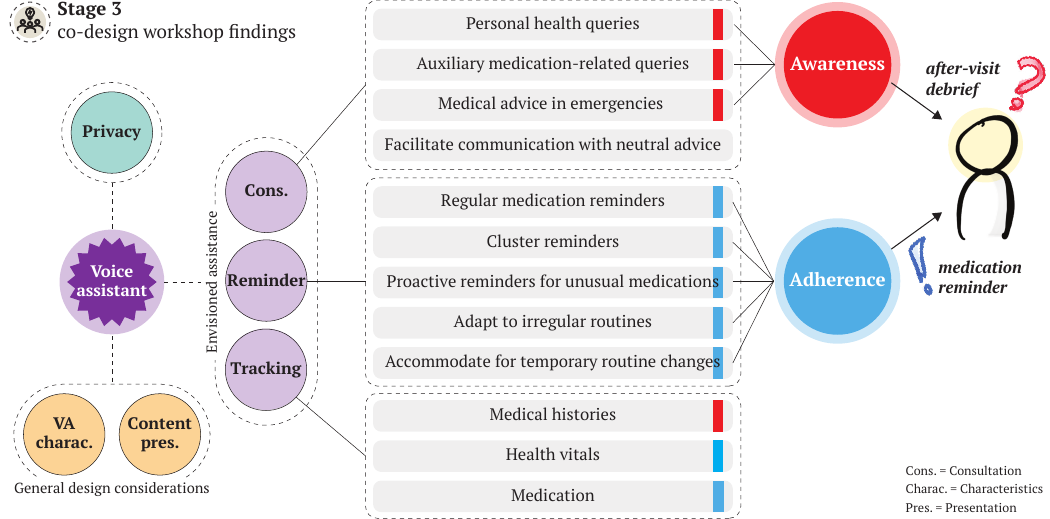}    
     \caption{Findings of co-design workshop for consultation, reminder and tracking. Our prototype does not focus on tracking. }
     \Description{This figures shows findings of the co-design workshop. On the left the envision assistance is shown with three themes: consultation, reminder and tracking. The voice assistant also has other findings outside of these three themes that are privacy and general design considerations (VA characteristics and content presentation). On the left side of the  figure the benefits to older patients are highlighted as awareness and adherence which are achieved through after-visit debrief and medication reminders respectively in our prototype.
     From the three themes the co-design findings are presented and then liked to the benefits for older adults: 
     1. Consultation: Personal Health Queries, Auxiliary Medication-Related Queries, Medical Advice in Emergencies, and Facilitate Communication with Neutral Advice. The first three of these are connected to Awareness.
     2. Reminder: Regular Medication Reminders, Proactive Reminders for Unusual Medications, Cluster Reminders, Adapt to Irregular Routines, and Accommodate for Temporary Routine Changes. All these are connected to Adherence.
     3. Tracking: Medical histories, health vitals, and medications. Medical histories is connected to labelled as awareness and health vitals and medications are labelled as adherence. But these are not liked to the adherence and awareness physically since our implementation does not include tracking. 
}
    \label{fig:co-design-findings-features}
\end{figure*}

\subsection{Findings: Consultation}
We identified four themes under consultation (Fig. \ref{fig:co-design-findings-features} and Table \ref{tab:findings-co-design}):

\subsubsection{Personal health queries}
During the \textit{after-visit debrief} with the prototype VA \highlight{lightgreen}{Phase 2}, we noted participants' questions, revealing their interests in specific health information. They asked questions related to the doctor's visit such as changes to medication doses and life style recommendations. The participants appreciated and found benefit in the VA's capability to facilitate personalized Q/A about their health during the after-visit debrief.

\subsubsection{Auxiliary medication-related queries}
Participants discussed how the extensive information provided with medications, such as side effects and interactions with other medications or foods, is often overwhelmingly complex and thus largely ignored ($n=2$). They suggested that VAs could alleviate this cognitive burden by summarizing and extracting key details about medications. 
Another issue raised involved the inadequate communication by doctors regarding the side effects of medications and their potential interactions with other health conditions ($n=3$). 
W2-L shared her frustrations: \pquotes{``Usually, they kind of explain it; they say okay, I am gonna change you from this medicine to that medicine because of this, and I am like fine. But then you actually get the medicine, and they give you a three-page thing with it. It's just easier sometimes to ask Alexa—like, tell me, I wanna know, can this medicine really harm me? Because usually, the doctor will tell you, I am changing you because of this, but they don't go through and [warm to] watch out for these things. For instance, one of the medications they put me on, it [Alexa] told me that my blood sugar may rise, so I was actually looking for it to rise, while the doctor only told me: This medication is better for [some condition]\dots 
but then Alexa told me my blood sugar may rise, and I was like, I don’t want my kidneys to go bad, but I don’t want my blood sugar to go up either. So, kind of okay, drop the ice cream because the medicine's gonna make your blood sugar go up.''} 
Such gaps in doctor-patient communication can leave patients confused and frustrated. VAs can serve as a secondary source of information, as W2-L noted, providing details on side effects that may impact existing conditions. W2-L also preferred using Alexa for medication information over traditional methods (healthcare portals and internet search) due to the convenience of speech interaction. Additionally, the VA can suggest actions to manage side effects, as seen when it motivated W2-L to make lifestyle changes by simply providing relevant information.
\subsubsection{Medical advice in emergencies}
Participants explored potential of VAs to act as surrogates for quick medical advice during emergency situations, especially, when immediate access to a doctor is unavailable ($n=3$). 
For instance, consulting a VA can be convenient for an initial assessment if someone misses a dose  or confuses their medications, as W1-P mentioned \pquotes{``I accidentally took XYZ\dots `not to worry' that sort of thing. You know sometimes you cannot get to your doctor.''} They also recognized that they can consult the VA during event-driven self-diagnosis, similar to typical internet searches, W1-B mentioned, \pquotes{``Can be as simple as Alexa, I have a bruise on my hand why is that. General medical advice\dots''}

\subsubsection{Facilitate communication with neutral advice} 
Participants discussed that VAs can be consulted as neutral third parties to facilitate communication between spouses or family members ($n=4$). For instance, W1-R suggested that Alexa could resolve disputes between caregivers by providing neutral, unbiased, yet tailored patient medical information. \sally{This finding highlights that while personal health assistants should provide contextualized advice, it should remain objective and factual.}

\begin{table*}[tb]
\centering
\caption{Summary of findings of co-design workshop for the envisioned personal health assistant: consultation, reminder, and tracking. The purple circles represent considerations that were implemented in the refined prototype through prompt engineering (Stage 4). }
\label{tab:findings-co-design}
\begin{tabular}{c p{8.1cm} p{8.2cm}}

&  \textbf{Challenges} & \textbf{Solutions} \\
\midrule
\midrule
\multirow{7}{*}{\rotatebox[origin=c]{90}{\textbf{Consultation}}}
& 1. Lack of personalized health queries in current VAs & Enable personalized health queries through integration of after-visit summaries from doctor's office \circlewithnumber{custompurple}{1}{1}\\
& 2. Information on medications is complex and doctors' may fail to inform important aspects & Enable auxiliary medication-related queries \circlewithnumber{custompurple}{2}{1}\\
& 3. Lack of immediate access to doctors in emergencies & Provide surrogate medical advice in emergencies \\
& 4. Conflict between patients and caregivers & Facilitate communication with neutral advice grounded in medical data \\

\midrule
\midrule

\multirow{12}{*}{\rotatebox[origin=c]{90}{\textbf{Reminder}}} 
& 1. Medication management is difficult: having to manage multiple medications, uncertain timing for taking medications, and forgetfulness  & Provide regular medication reminders as a back up to their current recall strategies  \circlewithnumber{custompurple}{3}{1}\\
&  2. Think of pills not as standalone items, but in relation to their intake time.  Additionally, may use pill boxes & Cluster medications by intake time for creating reminders. Consider their use of pill boxes \circlewithnumber{custompurple}{4}{1}\\
& 3. Missing out-of-routine medications (\eg allergy medications) is more common & Proactively remind about unusual and event driven medications\\
& 4. Irregular user routines; fixed reminders may not work & Capture and adapt to irregular user routine \circlewithnumber{custompurple}{5}{1}\\
& 5. Forget medications when travelling or away house & Accommodate for temporary routine changes \\
\cline{2-3}
\cdashline{2-3}
& \textit{User preferences for design alternatives} (finding of a direct question asked from participants) & Participants with many medications and varied routines preferred Design 2 (user-driven, tailored), while those with fewer medications preferred Design 3 (VA-driven). Create user profile \circlewithnumber{custompurple}{6}{1}\\

\midrule
\midrule

\multirow{6}{*}{\rotatebox[origin=c]{90}{\textbf{Tracking}}} 
& 1. Difficulties in recalling medical histories when needed & Track and provide medical history over time through after-visit summaries from doctors \\
& 2. Inconsistency and inaccuracy in tracking health vitals & Record, double check and report health vitals\\
& 1. Can miss medications despite reminders & Track medication intake to enable proactive actions \\
& 2. Unhappy with reliance on pharmacies for medication refill reminders & Track medication unavailability to alert for refills \\

\hline
\end{tabular}
\end{table*}

\subsection{Findings: Reminder}
\label{sec:co-design-findings-medication-reminders}
Below we discuss various possible breakdowns and their solutions in creating medication reminders \highlight{lightblue}{Phase 3}. The findings are summarized in Fig. \ref{fig:co-design-findings-features} and Table \ref{tab:findings-co-design}. 

\subsubsection{Support regular medication reminders} 
Some participants ($n=3$)
explicitly mentioned their need for medication reminders while others expressed their openness to it. The reasons identified for requiring medication reminders include: 1) managing multiple medications ($n=1$), 
2) uncertain timing for taking medications, often leading to irregular intake ($n=2$), and 3) the likelihood of forgetting to take medications when engaged in other tasks ($n=3$). These align with our interview (stage 1) findings. 
As a solution, participants discussed how having regular reminders could serve as back up and effectively support their existing medication management strategies ($n=7$). Additionally, participants discussed that reminders might need to become more intrusive, repetitive, and detailed upon cognitive decline ($n=4$).

\subsubsection{Cluster medications} 
While creating reminders with the prototype VA, we found that participants often preferred grouped reminders for morning and evening medications ($n=7$). Reasons included: 1) convenience of pill boxes ($n=5$), 2) cumbersome nature of setting individual reminders ($n=2$), and 3) benefits for people with disabilities---as noted by W1-B regarding his visually impaired friend, their medications are grouped and pre-packaged according to intake times.
Therefore, they wanted VAs to group medications into single reminders for specific times of or events during the day.  

\subsubsection{Reminders for unusual and event driven medications} 
Participants noted challenges with medications that require unusual timing or frequency ($n=2$), such as those taken every other day (W1-P) or in the evening outside the regular morning and night schedules (W1-L). For example, W1-L reported disrupted sleep due to delayed medication intake, which she mitigates by using Alexa to provide repeated reminders until she confirms intake. This underscores the need for precise and assertive reminders for  irregular medication schedules.
Additionally, some medications are event-driven ($n=1$). For example, W1-P highlighted the need for allergy medication in response to high pollen counts, which she is usually unaware of, \pquotes{``Alexa could say: Pollen count is high today. You should take some Claritin\dots cause half the time I don't realize until I'm already choking.''} 
This finding highlights the potential for VAs to proactively notify users about environmental factors impacting their health and the necessity for certain medications. 

\subsubsection{Capture and adapt to irregular user routine}
A potential breakdown identified is that some people might not have a consistent ``routine'' at all so creating regular reminders would be challenging.
Different solutions were discussed to address this challenge: 
1) implementing persistent reminders that repeat until the user confirms medication intake as evident from W1-L personal experience with unusual medication times (\pquotes{``To give you a possible solution, not only can I tell Alexa when it is a certain time for me to take a medication\dots
if I don't tell it I took my medication to remind me again in an hour\dots 
It's gonna keep reminding me every hour, until I take my medication.''}), 
2) reinforcing a better routine with wake-up and bedtime reminders, particularly for medications that must be taken at specific times, and 
3) designing adaptable reminders that adjust based on the user's reported wake-up time each day (most favored). 

\subsubsection{Accommodate for temporary routine changes} 
Participants highlighted forgetting medications while traveling, suggesting VAs should synchronize with personal calendars for packing reminders and adjust to time zones, sending notifications to mobile phones as smart speakers may not always be available ($n=4$). This need for adaptability was evident during interactions with the VA prototype, where participants ($n=3$) asked about its functionality across locations and automatic time zone adjustments. Concerns were also raised about handling minor \textit{one-off} routine changes, such as lunch outside or staying out late ($n=2$). 
They proposed three solutions: 1) integrating the personal calendar with the VA to remind them to take medications before commitments, 2) allowing the VA to adapt reminders for different daily schedules, and 3) enabling user-initiated conversations with the VA to update one-time schedule changes and adjust reminders accordingly.

At the end of the workshop, participants shared their preferences for medication reminder designs (Phase 3). Most favored Designs 2 and 3 over Design 1, seeing them as more ``personalized'' ($n=8$). W1-R and W2-L preferred Design 3 for its focus on gathering only essential information. W2-L valued the efficiency, stating, \pquotes{``Relevant questions. Just what do you need to know to help me do what I need to do.''} However, the majority in the first two workshops ($n=5$) favored Design 2 for its practical approach in building a user profile once before addressing medications.
Preferences reflected participants' medication management needs. For example, W1-R and W3-B, who take fewer medications, preferred the simpler Design 3, while W1-P and W1-L, managing multiple medications, found the comprehensive routine setup in Design 2 more useful and adaptable. 
All participants emphasized the importance of a flexible, context-aware VA (W1-B: \pquotes{``Certainly, it should be oriented to time and place''}). W2-N and W2-Y, who faced language barriers, could not establish a preference but acknowledged the system's benefits for older adults living alone. W3-Y noted, \pquotes{``I feel like this design was pretty good, especially for a single person, a single elderly living alone\dots it is very easy to forget and mix things up. So, it can remind me---even if it just said one sentence---and I would think: `Oh, I forgot to take my medicine. Let me go take my medicine quickly.' It would be a great help.''}

\subsection{Findings: Health Tracking}
Participants discussed various considerations for the envisioned VA to support health tracking (see Fig. \ref{fig:co-design-findings-features} and Table \ref{tab:findings-co-design}) 
\subsubsection{Track medical history}
Participants identified difficulties in recalling medical histories accurately ($n=4$), especially changes in medication dosage, when needed by existing and new doctors or emergency personnel. Currently, they track their medical history by reading after-visit summaries when needed ($n=2$), maintaining paper medication lists ($n=1$), or relying on their doctors to communicate updated lists with each other ($n=2$), but not to their satisfaction. They suggested the VA could help track and communicate updated medications with doctors directly or allow users to print ($n=2$). Tracking health information via VAs was also seen as beneficial for sharing data with family or friends ($n=5$). 

\subsubsection{Track vitals} 
Participants discussed the challenges they face and the utility of VAs in monitoring vitals \eg blood sugar and blood pressure ($n=4$). They noted that VAs could remind them to perform these checks per doctor’s instructions and prompt them to report readings. 
W1-P highlighted this benefit, \pquotes{``For a while, I had to take blood pressure, it's not something of a routine and that was a pain in the neck\dots
But Alexa could help me with the temporary program.''} 
Currently, participants track health metrics using traditional methods, such as writing them down or entering them into a health portal, which can lead to forgetting or inaccurate reporting. VAs can enhance this process, ensuring more accurate and consistent monitoring.

\subsubsection{Medication tracking} 
\label{sec:co-design-findings-medication-tracking}
After interacting with VA prototypes, participants also discussed effective ways to track medication intake ($n=8$). 
While open to tracking intake, preferences varied on the ``how''. Some preferred self-reporting (user-initiated) rather than VA-initiated inquiries. 
Conversely, some participants wanted more active monitoring methods, such as smart pill boxes and touch sensors. W1-R suggested these devices could communicate with the VA to tailor reminders, role-playing a scenario: \pquotes{``Hey, did you take the pill? Because the pillbox isn't touched.''}  
Similarly, W3-B wanted reminders if she left the house without touching the medication bottle. Overall, participants favored customizable interaction modes (user-initiated vs. VA-initiated) based on individual preferences.

Participants also expressed their dissatisfaction with reliance on pharmacies for medication refill reminders ($n=5$), as W2-M explained, \pquotes{``I get my medications currently from [pharmacy] online, and they send me text reminders, but they are not good at syncing the number of pills I have with when they prompt me to reorder. Sometimes, I almost have a full bottle left, and they ask if I want another. When I decline, they fail to prompt me again, so I have to go online.''}
Participants suggested that VAs could effectively remind them when refills are needed ($n=4$), as W1-P indicated, \pquotes{``At some point, I would like a little preview, maybe 5 days before, if I need to refill my medication.''}

\begin{figure*}[t]
     \includegraphics[width=\textwidth]{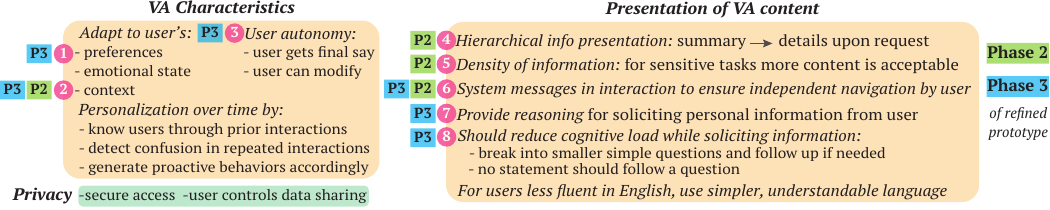}    
     \caption{Co-design workshop findings: general considerations for VA design. The pink circles represent considerations that were implemented in the refined prototype's phase 2 and 3 through prompt engineering.}
    \Description{The figure shows the general design considerations from workshop findings. It is mainly a textual figure. The considerations are marked with numbers: number 1 to number 8 to show what is implemented in the refined VA in next stage. P2 stands for phase 2 and P3 stands for Phase 3. The left side of the figure shows VA characteristics: Adapts to user's preferences (number 1, implemented in P3), emotional state, and context (number 2, implemented in P2 and P3). User autonomy (number 3, implemented in P3): user gets final say and user can modify. Personalization over time by: know users through prior interaction, detect confusion in repeated interactions, and generate proactive behaviors accordingly. 
    The right side of the figure shows: Hierarchical info presentation: summary details upon request (number 4, implemented in P2). 
    Density of information: for sensitive tasks, more content is acceptable (number 5, implemented in P2). 
    System messages ensure independent navigation by user (number 6, implemented in P2 and P3). 
    Provide reasoning for soliciting personal information from user (number 8, implemented in P3). 
    Reduce cognitive load while soliciting information: Break into smaller, simple questions and follow up if needed and No statement should follow a question (number 8, implemented in P3). 
    For users less fluent in English, use  simpler, understandable language. 
    The bottom left side shows privacy that has two elements, secure access and user controls data sharing.}
    \label{fig:co-design-findings-considerations}
\end{figure*}

\subsection{Findings: General Considerations for VA Design}
The general considerations for designing effective voice assistance are summarized in Fig. \ref{fig:co-design-findings-considerations}. 

\subsubsection{Privacy}
To enable the above mentioned features for consultation, reminder and tracking, \textit{health data sharing} with the VA  is crucial. 
Participants raised concerns about health data being shared with VAs as it can be accessed easily by unwanted personnel (strangers, visitors or friends) through the speech interface. 
However, they indicated they would be comfortable uploading documents into the system as long as they have control over the information they share, the data is anonymous and the access is secured ($n=7$).  
To address these concerns, following solutions were discussed among participants and researchers: 1) using confidential activation phrases and security questions for authentication,
2) restricting medical information access to authorized users only via voice profiles feature, and
3) allowing users to opt in or out of information sharing.

\subsubsection{VA characteristics}
\label{sec:co-design-findings-VA-characteristics}
Participants highlighted various aspects of VA characteristics in the given context: \\
\textbf{Adaptable to user.}
Participants discussed VAs should not be static across people but adaptable to users ($n=4$). 
Three aspects were discussed: 1) their preferences, 2) their emotional state, and 3) context of interaction.  
User preferences, specifically when explicit, should be taken into account as is.
Furthermore, W1-B advocated for incorporating users' feedback into the VA's responses (\pquotes{``Alexa, I don’t like the way you are talking to me.''}) allowing users to express their preferences, enabling the VA to adjust its behavior accordingly. 
The emotional state of the user and the context of interactions are also crucial ($n=4$). For example, 
W1-P described a potential scenario: \pquotes{``If I mixed up my medications, see, I probably wouldn't know. And then I’d panic. So if Alexa could sort of bring me down if I did something wrong and say: P, not to worry, if you are experiencing blah blah blah then you should call 911.''} 
Similarly, cognitive decline may necessitate adjustments as  W1-B pointed out, \pquotes{``When you are failing cognitively, you might want a very sophisticated answer, or you might not be able to deal with a sophisticated answer.''}\\ 
\textbf{Personalization over time.}
VA style should adapt over time as user preferences evolve ($n=6$). Personalization over time was discussed in two aspects. First, updating the VA's interaction style based on prior interactions, as W1-R noted: \pquotes{``Right now, every person is getting the same response from Alexa, but in theory, Alexa should begin to know who you are\dots 
So she should adjust its own presentation so that it minimizes counter-questions from you.''} 
Second, detecting confusion and frustration in repeated interactions to generate proactive VA behaviors, specifically in context of cognitive decline, as W1-B suggested:  \pquotes{``I can see how that can build in terms of when you ask for medication reminders and then you ask: 'Alexa, did I take so-and-so?' If there were a lot of 'Did I take so-and-so,' there should be a way to take account of that or notice that.''} 
Adjustments in style and tone were also suggested. 
W2-L proposed changes in voice tone after repeated failed reminders: \pquotes{``Like the third time it tells me to take my medicine and I have not answered, it might be nice to change the voice\dots A more aware voice. [Role playing] L! HAVE you taken your medicine?''} \\
\textbf{Support user autonomy and control.} 
Participants wanted to maintain autonomy over VAs, as observed during the \textit{medication reminders creation} interactions. They negotiated the frequency and timing of medication intake, revealing potential conflicts between user preferences and VAs' suggestions based on doctor's orders ($n=2$). While the VA initially encouraged adherence to medical recommendations, it eventually complied with participants' preferences, providing necessary warnings. This ``respect'' for user decisions was appreciated, W1-R: \pquotes{``It is always asking is it okay to have this appointment at 8 o'clock, do you want to change it. That's really good, it always asks you to make the final decision.''}
Participants also reported positive perceptions of the VA after interacting with it ($n=2$), feeling it enabled them to manage their healthcare independently and reduced the need to consult doctors for minor inquiries. W1-P highlighted:  \pquotes{``One of the nice things is that it puts you in control, you are not calling the doctor asking questions, you are managing your life. That’s nice.''}
Participants also expressed a desire for the VA to be easily modifiable and adaptable to individual needs and expectations ($n=5$). W1-R emphasized:  \pquotes{``Everything that she does should be modifiable except for critical health things\dots''} 
Thus, to ensure desired autonomy and control of user over interactions, two actionable items emerged: 1) users should have the final say, and 2) VA actions should be modifiable by the user.

\subsubsection{Presentation of VA content} 
We outline participants' preferences for presentation of VA content: \\
\textbf{Hierarchical content structure.}
Participants appreciated the hierarchical information presentation by the prototype VA, where a summary was followed by detailed information upon request as W2-L described, \pquotes{``It was like, the initial part was like a summary but then because I did not tell it to stop, it kept going and I kept listening, you know, at a point I had an answer okay so I said Alexa stop.''} \\
\textbf{Verbosity.}
Participants acknowledged the VA's verbosity was necessary due to the sensitive nature of medication discussions ($n=2$). W2-L noted: \pquotes{``I guess because it was talking about medication, it had to say more\dots You have to give more information so it's gotta solicit more information.''} \\
\textbf{Clear and simple Q/A format.}
Multiple questions in a single prompt increased cognitive load. 
During the user profile creation, W2-M, who has a variable routine, was overwhelmed by the VA asking about meal times on both early and late wake-up days in the same question. W2-L suggested a solution: breaking down complex questions into simpler, sequential ones, such as, \pquotes{``I almost wanted it to say [Alexa:] Okay, what's your schedule for Monday? What's your schedule for Tuesday? \dots Kind of like how I'd actually put it on my calendar. ''} This would reduce cognitive load by tailoring further questions based on each response.
Additionally, participants were sometimes confused during questions. After asking a question, the VA would sometimes immediately begin an explanation, such as, \textit{``Let's start with your first Amoxicillin. Could you tell me your usual meal times? This will help me schedule your Amoxicillin reminders accurately\dots''} Participants often began answering right after the question, but the VA continued speaking, causing confusion.
They also found it unclear at times why the VA asked personal questions and expected it to communicate its intentions first. \\
\textbf{Simple, understandable language.} 
Language barriers were evident in immigrant participants' (W3-N and W3-Y) interactions with the VA, as they sometimes did not understand the VA's questions and statements and needed more time to mentally translate their responses from Mandarin to English. This indicates that, when designing for special populations within aging adults (\eg immigrants less fluent in English), considerations such as simpler phrases, shorter sentences, and less complex language should be employed. 

The workshops resulted in considerations for both the application-specific VA (Table \ref{tab:findings-co-design}) and general voice assistance (Fig. \ref{fig:co-design-findings-considerations}).
The next stage of our design process focuses on refining the prototype based on key considerations identified during the co-design workshops to better meet older adults' needs for a personal health assistant.

\section{Stage 4: Refining the Prototype VA}
\label{sec:refinement-VA}
The co-design workshop revealed several key design considerations (summarized in Fig. \ref{fig:co-design-findings-features} and Table \ref{tab:findings-co-design}) that are actionable within the envisioned personal health assistant to support awareness and adherence for older adults.
To refine the initial prototype, we focused on consultation and medication reminders, excluding tracking, as it cannot be validated in a single interaction. General design recommendations (Fig. \ref{fig:co-design-findings-considerations}) were applied to the VA where feasible, given the brief nature of the interaction for validation study. For example, long-term personalization cannot be assessed in one session. 
In this section, we detail out the implementation of the refined health assistant. 
The interaction with the VA involved three same phases (Fig. \ref{fig:example-interaction}) as initial prototype (stage 2). Purple circles represent considerations for the envisioned assistance and pink circles represent general considerations.
The three modified phases are as follows:

\subsection{\highlight{lightyellow}{Phase 1} Health Data Sharing}

Phase 1 targets capturing patient's health data. We simulate health data sharing with the prototype by having the users fill out their chronic conditions, medications and specialists they see. 
The medical data is then input into an LLM prompt by the experimenter to convert it into an after-visit summary following a real summary template (see Fig. \ref{fig:example-interaction}).

\subsection{\highlight{lightgreen}{Phase 2} After-Visit Debrief}
The after-visit debrief interaction is the same as the initial prototype (section \ref{sec:prototype1-summary}). Fig. \ref{fig:example-interaction} shows the system prompt to support hierarchical presentation \circlewithnumber{custompink}{4}{1} of dense information \circlewithnumber{custompink}{5}{1}. The VA allows personal health \circlewithnumber{custompurple}{1}{1} and auxiliary medication queries \circlewithnumber{custompurple}{2}{1} grounded in the users' health data \circlewithnumber{custompink}{2}{1}. Once the user is done with this phase, the VA guides the user to phase 3 \circlewithnumber{custompink}{6}{1} by saying: \vquotes{``Great, let's move on to the next task. I need to know a bit about your daily routine and how you manage your medications to suggest appropriate reminders. Let me know when you are ready.''}

\subsection{\highlight{lightblue}{Phase 3} Medication Reminder Creation}
We implement the VA to support adherence by enabling users to create regular medication reminders \circlewithnumber{custompurple}{3}{1}. The VA should accommodate for variations in the user’s routine to ensure that the reminders are timely and relevant \circlewithnumber{custompurple}{5}{1}. Designing such an adaptable VA can be achieved by 
capturing user routine---relevant to their medication management---to create a user profile \circlewithnumber{custompurple}{6}{1}. 
The medication reminder creation consists of two sub-phases:
\subsubsection{Creating user profile}
In this sub-phase, the VA focuses on gathering requisite information about the user's medication routine and preferences to create a profile \circlewithnumber{custompurple}{6}{1}. The AVS is input into the LLM to extract medication requirements as prescribed by doctors. 
In co-design workshops, most participants ($n=5$) preferred Design 2 (user profile creation before reminders) for its comprehensive approach, which is helpful for those with multiple medications or varying routines. However, participants with fewer medications ($n=3$) preferred Design 3 (asking only relevant questions). To accommodate the majority, we implemented a hybrid of Designs 2 and 3---creating user profile first but only asking essential questions.
To create user profile, we prompt (Fig. \ref{fig:example-interaction}) LLM to ask the following: 
\begin{enumerate}
 [leftmargin=*, nolistsep]
    \item \textbf{Assess routine consistency:} To account for irregular schedules for users, the VA first asks about the consistency of the user's routine, including variations on weekends \circlewithnumber{custompurple}{6}{1}. 
    \item \textbf{Current medication management:} 
    The VA asks the user how they manage their medications, whether the user organizes the medications 
    into pill boxes by time of day (\eg morning and night pills) or use  original containers \circlewithnumber{custompink}{2}{1}. 
    \item \textbf{Need-based queries:} 
    The VA asks necessary questions about their routine based on the medication list in the AVS. For instance, the VA only inquires about meal times if a medication needs to be taken around meals.
    These questions are meant to capture their daily routine, meal times, availability for medication intake, and any other specific preferences for reminders. 
\end{enumerate}

After completing the questions, the LLM component generates a user profile (Fig. \ref{fig:example-interaction}). The VA then proceeds, saying, \vquotes{``Let's create medication reminders now. Let me know when you're ready.''}

\subsubsection{Grouping medications for suggesting reminders}
\label{sec:final-prototype-clustering-medications}
The next step involves organizing medications and suggesting reminders tailored to the user profile as a backup to current management practices \circlewithnumber{custompurple}{3}{1}. The LLM uses the AVS and the user profile as input 
to suggest reminders in chronological order, following the rules below:

\begin{itemize} [leftmargin=*, nolistsep]
    \item \textbf{Cluster medications:}
    The VA groups medications by time of day (morning, afternoon, night) or event (meals, wake-up, sleep), while also considering pill box use \circlewithnumber{custompurple}{4}{1}. For example, bedtime medications are grouped as one reminder: \vquotes{''X and Y should be taken at bedtime. I'll set the reminder for 9:45 p.m., just before bed. Does that sound good?''} 
    \item \textbf{Adapt to the user:} The VA uses the user profile to suggest reminders tailored to the user's routine and practices \circlewithnumber{custompink}{2}{1}. It integrates medications into the daily schedule and suggests reminders in chronological order. For example: \vquotes{''Let's start with your morning medication\dots Next, the afternoon medications\dots''} 
    The VA also reads out the name of each medication in a grouped reminder to ensure consistency with the user's pill management. For instance, it checks if the medications in the evening reminder match those in the evening pillbox: 
    \vquotes{''We'll group your evening medications---A, B, and the second dose of C together\dots''}

    \item \textbf{Medication intake compliance vs. user control:}  
    The VA prioritizes medication intake requirements from the AVS when suggesting reminders, while allowing the user to make the final decision \circlewithnumber{custompink}{3}{1}. However, the VA emphasizes following the doctor's instructions and alerts the user if their request conflicts with these requirements, as allowing the user to proceed with their preference may not always be beneficial. A balance between user autonomy and medical compliance is necessary.
    For example, the VA might say: \vquotes{``Setting a reminder for X at 11 AM conflicts with the doctor's instructions to take it in the morning before food or other medications. Would you like to keep the 8:15 AM reminder, or proceed with 11 AM?''}  
    \item \textbf{Modifiable reminders:} The VA can update, add, or delete reminders upon user request at any time \circlewithnumber{custompink}{3}{1} to match their preferences \circlewithnumber{custompink}{1}{1} and even adjust the frequency of any reminder \circlewithnumber{custompurple}{5}{1}. 
    
\end{itemize}
The VA lets the user know once all reminders are set and user can end the interaction by saying \vquotes{``Goodbye''}. 
\subsubsection{Implementation of General Considerations.}
Based on workshop findings, the following general considerations were implemented to improve user experience:
\begin{itemize} [leftmargin=*, nolistsep]
    \item \textbf{Single question approach:} The VA asks only one question at a time to avoid confusion and reduce cognitive load, even if multiple follow-up questions are needed \circlewithnumber{custompink}{8}{1}.
    Complicated follow-up questions are split into simple, single questions. For instance, \vquotes{``What time do you wake up and go to bed on weekdays?''} followed by \vquotes{``Now, how about weekends?''} if the user has a different routine on the weekend.  
    \item \textbf{Explain reasons behind questions:} Before each question requesting personal data, the VA explains its purpose to ensure clarity and transparency \circlewithnumber{custompink}{7}{1}. 
    For example, before asking about the user's routine, it says: \vquotes{``I need to know a bit about your routine and how you manage your medications to suggest appropriate reminders.''} 
    Additionally, no statements should follow a VA question to avoid confusion \circlewithnumber{custompink}{8}{1}.
    \item \textbf{Guided user navigation:}
    are added to ensure that users can navigate the various components of the interaction independently \circlewithnumber{custompink}{6}{1}.
    For instance, after completing the profile creation, the VA says, \vquotes{``Let's create medication reminders now. Let me know when you're ready.''}

\end{itemize}

\begin{figure*}[t]
     \includegraphics[ width=0.98\textwidth]{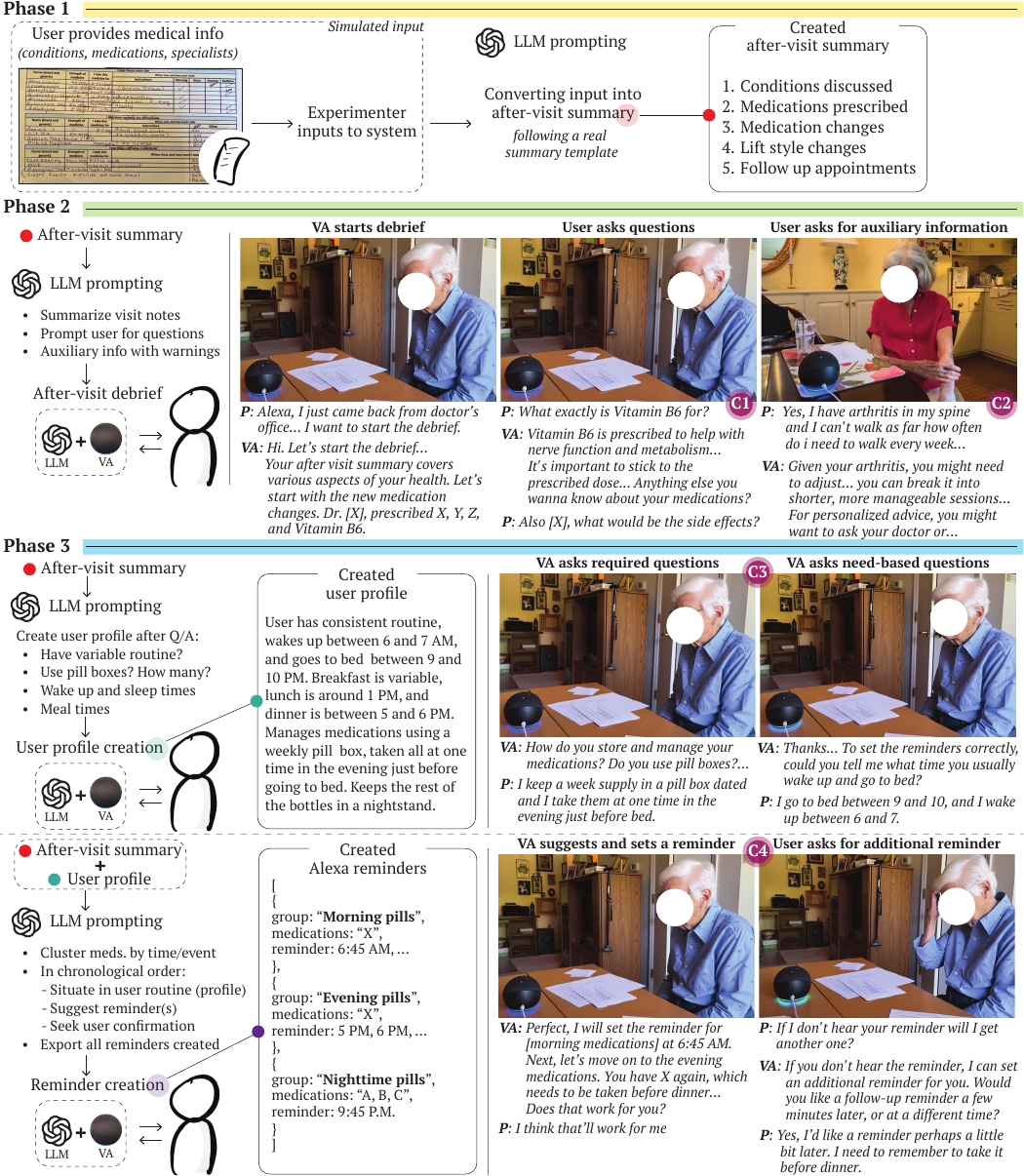}         \caption{Refined VA and example participant interaction \href{https://youtu.be/i2-Ewn7_Rjw}{[\underline{interaction video}]}. ``\dots'' represents truncated speech for brevity. C1--C4 are conversation labels used in the text to reference examples in this figure. All prompts in supplementary materials$^1$.} 
     \Description{This figure shows an interaction flow with system details and examples. It shows three phases of interaction:
     Phase 1: Initial Input and AVS Creation:
     The user provides medical information, including conditions, medications, and specialists. The experimenter inputs this information into the system. LLM prompt is: ``convert input into AVS following a real summary template"  The created AVS includes: Conditions discussed, Medications prescribed, Medication changes, Lifestyle changes, and Follow-up appointments. 
     Phase 2: Phase 2 has two parts, First, the created AVS is input to an LLM prompt: ``Summarize visit notes. Prompt user for questions. Auxiliary info with warnings.'' After-visit debrief happens between the LLM-powered VA and the user (older adult). Example interactions are:
    1. VA starts debrief. P: Alexa, I just came back from doctor’s office... I want to start the debrief. VA: Hi. Let’s start the debrief... Your after visit summary covers various aspects of your health. Let’s start with the new medication changes. Dr. [X], prescribed X, Y, Z, and Vitamin B6.
    2. User asks questions (conversation labeled as C1). P: What exactly is Vitamin B6 for? VA: Vitamin B6 is prescribed to help with nerve function and metabolism... It's important to stick to the prescribed dose... Anything else you wanna know about your medications? P: Also [X], what would be the side effects?
    3. User asks for auxiliary information (conversation labeled as C2): P: Yes, I have arthritis in my spine and I can't walk as far how often do i need to walk every week... VA: Given your arthritis, you might need to adjust... you can break it into shorter, more manageable sessions... For personalized advice, you might want to ask your doctor or...
    Phase 3: AVS is input to an LLM prompt: ``Create user profile after Q/A: have variable routine? use pill boxes? how many? wake up and sleep times? meal times? The user-VA interaction driven by this prompt results in created user profile such as: ``User has consistent routine, wakes up between 6 and 7 AM, and goes to bed between 9 and 10 PM. Breakfast is variable, lunch is around 1 PM, and dinner is between 5 and 6 PM. Manages medications using a weekly pill box, taken all at one time in the evening just before going to bed. Keeps the rest of the bottles in a nightstand.''
    Example interactions for the first part of phase 2 are (these are labelled as C3): 
    VA asks required questions: VA: How do you store and manage your medications? Do you use pill boxes?... P: I keep a week supply in a pill box dated  and I take them at one time in the evening just before bed.
    VA asks need-based questions: VA: Thanks for sharing that. To set the reminders correctly, could you tell me what time you usually wake up and go to bed?
    P: I go to bed between 9 and 10, and I wake up between 6 and 7.
    Second part of phase 2 starts with AVS and user profile input into LLM prompt: `` Cluster meds. by time/event. In chronological order: situate in user routine (profile), suggest reminder(s), seek user confirmation. Export all reminders created. The user-VA interaction driven by this prompt results in created Alexa reminders in json format for Alexa to set them: [
    {
    group: “Morning pills”,
    medications: “X”,
    reminder: 6:45 AM, ...
    },
    {
    group: “Evening pills”,
    medications: “X”,
    reminder: 5 PM, 6 PM, ...
    },
    {
    group: “Nighttime pills”,
    medications: “A, B, C”,
    reminder: 9:45 P.M.
    }
    ]
    Example interactions are (labelled as C4): 
    1. VA suggest and sets a reminder: VA: Perfect, I will set the reminder for [morning medications] at 6:45 AM. Next, let’s move on to the evening medications. You have X again, which needs to be taken before dinner... Does that work for you? P: I think that’ll work for me.
    2. User asks for additional reminder: P: If I don't hear your reminder will I get another one? VA: If you don't hear the reminder, I can set an additional reminder for you. Would you like a follow-up reminder a few minutes later, or at a different time?
    P: Yes, I’d like a reminder perhaps a little bit later. I need to remember to take it before dinner.
    }
    \label{fig:example-interaction}
\end{figure*}
\section{Stage 5: In-Home Validation Study}
\label{sec:evaluation-study}

To validate the second iteration of the personal health assistant, we conducted a small-scale in-home study with older adults, assessing the VA's usability in debriefing after-visit summary and guiding users to create reminders.

\subsection{Study Plan and Procedure}

At the beginning of the study, each participant received a brief description explaining that they would be interacting with a VA designed to assist with their health self-management. Participation was voluntary, with consent obtained via signed forms. Participants then completed a medical information form, detailing their medical conditions, medications, and the specialists they usually see. The experimenter informed them that this data (anonymous) would be used to simulate the AVS for interaction with the VA. While the experimenter set up the task, participants filled out a demographics survey.
The experimenter then explained the interaction and demonstrated a few exchanges with the VA. 
Participants could ask questions they had before they began interacting with the VA independently. 
\sally{The system guided the user through the process of AVS debrief (Phase 2), user profile creation (Phase 3), and reminder creation (Phase 3) (see \href{https://youtu.be/i2-Ewn7_Rjw}{\underline{interaction video}}). Participants were able to freely engage with the system for as long as they wished, \ie they were able to ask as many questions to the VA as they wished.}
After the interaction, participants were asked to fill out a usability survey. Finally, the study ended with a brief interview to gather feedback. 
The entire study took approximately 60 minutes.

\subsection{Participants}
We recruited five older adults for the validation study (4 female, 1 male), aged 65 and above $(M=78.80$, $ SD=4.76$). Participants included a mix of those familiar with VAs and those who were not, as well as individuals who had previously participated in various stages of the process (interviews and co-design workshops) and those who had not (Table \ref{tab:evaluation-study}).

\subsection{Data Analysis and Metrics}
\sally{We evaluated the usability of the system through analyzing the audio interaction data and the System Usability Scale (SUS) score from the post-study questionnaire}: 
\subsubsection{User interaction evaluation} 
We coded interactions with the VA during different stages to evaluate its \sally{usability}. \amh{We matched participant conversations with the VA's responses, as well as its generated personas and reminders, to determine whether the VA fulfilled its intended functionality.}
\begin{itemize} [leftmargin=*, nolistsep]
    \item \textbf{After-visit debrief.} 
    We coded whether participants successfully \textit{navigated through the VA's debrief} \sally{and were debriefed on all aspects of the after-vision summary} and the \textit{percentage of questions successfully answered} \sally{by the VA}.
    \item \textbf{User profile creation.} 
    We coded \sally{\textit{how accurately the persona} generated reflected the user's routine based on their conversation with the VA.}
    \item \textbf{Reminder creation.} 
    We coded the \textit{percentage of reminders successfully created} during the interaction and \sally{ how \textit{accurately the created reminders} matched the user profile and their conversation with the VA.}
\end{itemize}

\subsubsection{\sally{Usability score }}
We calculated the System Usability Scale (SUS) score for each participant using a 10-item, five-point Likert scale questionnaire \cite{brooke1996sus}. A score above 70 ($0-100$) indicates good system usability \cite{bangor2009determining}. We also gathered participants' \sally{perceptions of the usability of the VA and whether they would use the VA again} during post-study interviews to complement the usability score.

\begin{table*}[t]
\centering
\caption{Findings of Validation Study: \textbf{Prior VA experience} indicates participants' familiarity with the VA and their participation in various stages of the design process. \textbf{Debrief} shows whether participants successfully navigated the after-visit debrief with the VA. \textbf{Reminders Created} shows the number and percentage of medication reminders successfully created, e.g., 4/4 (100\%), and $r=3$ indicates that 3 reminders were created for morning, evening, and bedtime.  
\textbf{*}\textit{All reminders created were accurate.}
}
\label{tab:evaluation-study}
\begin{tabular}{l | >{\raggedright\arraybackslash}p{1.5cm} >{\raggedright\arraybackslash}p{2cm} >{\raggedright\arraybackslash}p{2cm} | c c >{\raggedright\arraybackslash}p{1.8cm} >{\centering\arraybackslash}p{1.2cm} >{\raggedright\arraybackslash}p{4cm}}  

\multicolumn{4}{r}{\textbf{Medication Management and VA use}} & \multicolumn{5}{c}{\textbf{Results of Validation Study}} \\
\toprule
\textbf{P} & \textbf{Pill boxes} & \textbf{Medications (doses)} & \textbf{Prior VA experience} & \textbf{SUS} & \textbf{Debrief} & \textbf{Q/A during debrief} & \textbf{User profile creation} & \textbf{Reminders created$^{*}$}  \\ 
\midrule
\midrule
1C & evening & 4 (5 doses) & novice to VAs & 95.0 & \checkmark & 100\% \newline 2/2 answered & \checkmark & 4/4; 100\%; $r=3$ \newline morning, evening \& bedtime  \\ 
2C & morning \& bedtime & 7 (5 doses, 2 as needed) & novice to VAs & 87.5 & \checkmark & 75\% \newline 3/4 answered & \checkmark & 5/5; 100\%; $r=3$ \newline morning, bedtime \& standalone  \\ 
3L & morning \& bedtime & 15 (18 doses, 1 as needed) & workshop only, novice to VAs & 62.5 & \checkmark & 100\% \newline 3/3 answered & \checkmark  & 13/14; 92.86\%; 3 \newline morning (2) \& bedtime \\ 
4M & $\times$ & 5 (6 doses) & whole pipeline  & 90.0 & \checkmark & 100\% \newline 4/4 answered & \checkmark & 5/5; 100\%; $r=3$ \newline morning, evening \& night \\ 
5R & $\times$ & 2 (2 doses) & whole pipeline  & 90.0 & \checkmark & 100\% \newline 1/1 answered & \checkmark & 2/2; 100\%; $r=1$ \newline bedtime  \\ 

\hline
\end{tabular}
\end{table*}
\subsection{Results} 
The results of validation study are summarized in Table \ref{tab:evaluation-study}. 
\subsubsection{User interaction evaluation} 
Our results on the VA's effectiveness, as reflected in user interactions, are as follows:\\
\textbf{After-visit debrief.} 
All participants ($N=5$) successfully navigated the AVS and engaged with the VA, each asking at least one question, with an average of three per participant. They asked a variety of questions about their health \eg medical conditions (C1  in Fig. \ref{fig:example-interaction}) and lifestyle changes (C2). All questions were answered by the VA, except one that was not heard, demonstrating the ability of the VA in handling personal health queries. The VA also personalized responses based on the user's health context from the AVS. For example, it tailored advice to the participant’s arthritis (C2).\\
\textbf{User profile creation.} Profiles created for all the participants ($N=5$) were accurate according to their conversation with the VA. Fig. \ref{fig:example-interaction} shows an example of created user profile. \\
\textbf{Reminder creation.} 
The VA successfully created reminders for all medications listed in the AVS, except for one medication. Details of the created reminders are shown in Table \ref{tab:evaluation-study}. Medications were grouped into $1-3$ reminders based on the time of day or events. All reminders were accurate as per the user profile, medication requirements, and preferences recorded during the conversation. These results show the VA effectively adapted to user routines when suggesting reminders. Additionally, users easily modified reminder timing and frequency (C4 in Fig. \ref{fig:example-interaction}).

\subsubsection{Usability evaluation}

The average SUS score was 85 ($SD = 12.90$), which is above the 70 threshold for good (acceptable) usability \cite{bangor2009determining}, with only one participant scoring below \amh{80 (excellent/ acceptable)}. \amh{Participant 3L rated the system as ``ok'' or ``marginal'' (62.5/100), explaining that she currently does not find a need for reminders---consistent with prior research a small portion of users (9\%) were not interested in reminders for medications \cite{pater2017addressing}. }

\noindent \textbf{User perceptions. } Participants found the system learnable, as 4M noted, 
\pquotes{``Just a matter of doing it and practicing and getting used to it.''} Even novice participants found the system intuitive. For instance, participant 1C found the system easy to use and navigate despite her preconceived notion of a low 
learning curve with technology prior to this interaction: \pquotes{``The experience was very pleasant, understandable, to grasp 
\dots I caught on very quick and I think the more I used it, the better I would get. I just, I don't have one here and it was suggested that I don't get too used to something like this. So, I, we have avoided it.''}
However, participants with low tech familiarity (\eg non-smartphone users) may need assistance inputting the AVS, as 1C added, \pquotes{''I'd need a crash course on taking the picture of the medication.''} 
Our validation study showed that participants found the VA intuitive and easy to use; they successfully navigated the after-visit debrief and medication reminder creation. Overall, participants perceived the VA as learnable and pleasant, indicating that the design considerations (Figs. \ref{fig:co-design-findings-features} and \ref{fig:co-design-findings-considerations}) were effective in creating a fluid, positive interaction.

\section{General Discussion and Conclusion}

In this work, we tackle the lack of relevant user context (\ie user routine and health data) and usability challenges in VA technology to support older adults' health self-management. We co-designed an LLM-powered VA---personalized to their health information and aligned with their self-management practices---that \amh{successfully} debriefs doctor visits, facilitates personal health queries, and assists in creating medication reminders to address identified health \textit{awareness} and \textit{adherence} challenges. 
Below, we describe our implications for \amh{the design of VAs in health self-management and for enhancing user-agent interactions more broadly.}

\subsection{\amh{Design Implications for Improving VAs to Support Health Self-Management}}
\amh{Below, we present design implications for VAs to support awareness and adherence aspects of health self-management.} 

\subsubsection{\amh{Improving awareness: health awareness beyond personalized Q/A}}

\amh{We implemented personalized Q/A as a step towards enhancing user engagement with health data to ultimately improve awareness. During co-design workshops and validation studies, participants asked health-related questions and follow-ups, with the VA responding successfully without breakdowns (C1 and C2, Fig. \ref{fig:example-interaction}), overcoming a common limitation of commercial VAs \cite{mahmood2024situated, bickmore2018patient}.}

\amh{Older adults envisioned the proposed personal health assistant playing a broader role beyond after-visit debriefs and personalized Q/A. Findings from our co-design workshops suggest that they desire the VA to adapt its role based on the context, even if the core function remains medical information seeking. For example, they wanted the LLM-powered VA to act as a \textit{surrogate for quick medical advice} during emergencies, which is a step up from previously explored patient-provider communication that primarily focused on symptom collection and reporting to care providers \cite{yang2024talk2care, jo2023understanding}. Additionally, participants highlighted the VA's potential to serve as a \textit{neutral facilitator during medical decision-making}, particularly when receiving conflicting advice from multiple caregivers. By leveraging users’ medical histories from AVSs, the VA could empower users with relevant information to enable informed decision making.
Additionally, the VA can support self-advocacy---essential for older adults' well-being \cite{ruggiano2016if}, given their limited health literacy and the shortcomings of the healthcare system \cite{doyle2019managing, nguyen2022systematic}---by presenting factual information about treatment plans, medication changes, and related suggestions. As noted during initial interviews, specific aging groups such as assisted community residents reliant on nursing staff, may benefit greatly from self-advocacy support to address unsatisfactory care.}

\amh{While LLM-powered VAs hold significant potential for supporting health awareness, challenges remain with the reliability of LLMs in medical contexts. LLMs  can occasionally deliver contradictory, illogical, or even hallucinated responses \cite{koubaa2023exploring}, which can pose serious risks to users in medical scenarios \cite{ong2024ethical, hastings2024preventing}. Misinformation is also prevalent in commercial VAs that, if acted upon, could cause significant harm to users \cite{bickmore2018patient}. Thus, future work is needed to develop benchmarking frameworks, risk-assessment methodologies, supplementing training data, and necessary safeguards for employing LLM-powered VAs in health applications \cite{ong2024ethical, hastings2024preventing}. }

\subsubsection{\amh{Improving medication management: personalization through task and user context}}
\amh{VAs often fail to provide tailored support due to a lack of task-specific, contextually relevant information, leading to usability issues such as irrelevant details or confusing reminders \cite{mahmood2024situated}. Inputting medical data, such as medication timings, is frustrating and time-consuming for older adults, particularly during initial setup \cite{grindrod2014evaluating, jesus2020voice, pater2017addressing, deutsch2016smartwatch}, reflected in below-acceptable usability (SUS) scores for several e-medication platforms ($M=52.8$ across 21 applications) \cite{grindrod2014evaluating, patel2020prospective}.
Unlike text-based applications requiring manual entry, our system shifts responsibility to the VA by enabling users to upload images or PDFs of AVSs. Additionally, leveraging improved image-to-text models reduces errors in manual entry \cite{raj2023revolutionizing, arts2002defining}.}
\amh{In addition to medication timings \cite{grindrod2014evaluating, jesus2020voice, pater2017addressing, deutsch2016smartwatch}, the proposed system leverages other key information from AVSs such as medication requirements---necessary for setting reminders that align with doctor's orders as emphasized in prior work \cite{deutsch2016smartwatch}.  }

\amh{Lack of meaningful reminder messages to convey the purpose make them ineffective and unreliable, as users are often left wondering what the reminder is for \cite{mahmood2024situated, pater2017addressing}. This issue is particularly challenging for individuals with mild cognitive impairment (MCI) \cite{mathur2022collaborative}. Our findings further suggest that reminder messages should be personalized to align with users' medication intake practices and preferences. For example, participants preferred clustering medications by time of day or organizing reminders around their pillbox usage. Instead of listing medication names, they preferred prompts such as ``morning medications.'' To enhance usability, our system adds details to reminders about medication requirements or precautions. For instance, a reminder might say: \textit{``Please take your morning pills. Amoxicillin can upset the stomach; take it with food.''} (see VA prompt in supplementary materials\footnotemark[1]). 
Personalized messages with \textit{situated awareness}---contextually relevant instructions such as \textit{``take it with food''}---can help older adults manage multiple medications with specific requirements (\eg take before meals, on an empty stomach, or avoid certain foods), an otherwise challenging task, especially for those with MCI \cite{chen2021understanding}. While such messages are beneficial, their repetitiveness may negatively impact user experience.
The flexibility of our designed system allows users to customize reminder messages during creation. Future work should explore the optimal information density for these messages to better serve diverse aging populations.}

\amh{In addition to having medical information available, we found that incorporating user context and preferences is also critical for usable medication reminders. During co-design workshops, participants emphasized the importance of tailoring reminders to their routines (timing) and medication management practices (using pill boxes). Failure to adapt to user routines leads aging adults to discontinue using medication reminder technologies due to untimely notifications and over-reminding (often used to compensate for missed reminders) \cite{pater2017addressing}. Additionally, over-reminding MCI individuals can lead to over-medication \cite{mathur2022collaborative}. Aligning reminders with users' varying routines, achieved through creating user profiles in our designed VA, offers the potential to reduce missed reminders by ensuring timely notifications. Furthermore, our system avoids over-reminding by not repeating a reminder once the user has acknowledged it.}

\amh{Most participants preferred familiar methods and wanted VAs to adapt to their current practices, aligning with findings for individuals with MCI \cite{mathur2022collaborative}. Integrating solutions into existing medication management routines is crucial for adoption, as older adults may resist abandoning familiar practices. Proactive systems, such as smart pill boxes \cite{xu2016medhelp, suzuki2014smartphone, mugisha2017framework}, could address the lack of active medication tracking and tailor reminders for diverse cognitive and physical needs. For instance, they could reduce over-reminding by reminding users only if the pillbox remains unused, as noted by participant W1-R (Sec \ref{sec:co-design-findings-medication-tracking}).  However, it is pertinent to ensure that older adults' routines and autonomy are respected. Future work should focus on designing proactive VAs that adapt to users’ established practices and diverse needs.}

\subsection{\amh{Design Implications for Improving User-Agent Interactions}}

\amh{Below, we present design implications to address usability challenges in user-agent interactions.}

\subsubsection{\amh{Intuitive and learnable interactions}}

\amh{Older adults have different speaking styles, accents and fluency. Thus, they may need to do code-switching to converse with VAs \cite{harrington2022s}, not only in how they say it but also in what they choose to say \cite{mahmood2024situated, pradhan2020use, pradhan2018accessibility}.  However, our observations show that due to the LLM's contextual understanding and its ability to mitigate speech recognition and intent recognition errors \cite{mahmood2025user}, participants---regardless of their prior experience with VAs---found the system easy to use.
Unlike prior work \cite{mahmood2024situated}, our implementation of an LLM-powered VA introduced navigation cues to guide and support independent interaction. 
While the VA effectively guided participants within a single interaction, it struggled with providing guidance across interactions. For instance, when participant 1C asked how to access the personal health assistant later, the VA gave a generic response, \textit{``Feel free to reach out at any time. I am here,''} instead of a clear instruction, such as, \textit{``To contact me later, just say: Alexa, personal health assistant''}. To address this, the VA should provide actionable guidance on system use and capabilities both within and between interactions. Additionally, older adults with limited technological familiarity may require extra support for tasks such as entering medical information and initiating interactions with the VA. Step-by-step assistance, context-aware responses, dynamic tutorials, and real-time guidance are essential for enabling older adults to engage effectively with conversational VAs and achieve their goals \cite{mahmood2024situated, koon2020perceptions}.}
Additionally, older adults less familiar with technology may need extra support, such as entering medical information and starting interactions with the VA. The VA should offer step-by-step assistance for these tasks as well. Context-aware responses, dynamic tutorials, and real-time guidance are key to helping older adults engage effectively with conversational VAs to achieve their goal \cite{mahmood2024situated, koon2020perceptions}.

\subsubsection{\amh{Flexible interactions: adaptable and modifiable}}
\amh{ While prior work has demonstrated that older adults prefer linear navigation when setting medication reminders \cite{grindrod2014evaluating}, our observations revealed additional user needs. Participants generally appreciated guided navigation and found the system useful, but they often broke the conversational flow to ask various questions related to other aspects of medications, such as side effects and interactions, while setting reminders. They also negotiated the timing and frequency of reminders (C4, Fig. \ref{fig:example-interaction}). Thus, our findings indicate a need for flexibility in two key areas: 1) an adaptable conversational flow that includes interaction cues and guided interactions to support older adults, while allowing diversions within conversational units, and 2) modifiable functional features such as the ability to modify the timing and frequency of reminders.
Such flexibility in interactions assures user control and autonomy. However, excessive user control may conflict with user needs, specifically for medication adherence. Thus, there is a need for balance between user preferences and medical requirements. 
In our designed LLM-powered VA, such adaptability can be achieved through simple adjustments to LLM prompts---that does not require domain expertise. Options for users and caregivers to dynamically update the prompts---via voice commands or text interface---can enhance usability of the system for users with diverse needs. }

\subsection{Limitations and Future work}
While our five-stage design process offered valuable insights into older adults' needs and preferences for personal health assistants and validated the prototype's usability, a \amh{large-scale longitudinal field study is needed to assess the usability and effectiveness of our VA} to increase awareness of their health and enhance their compliance with medication reminders. Additionally, our small, sample---focused on one community center and local residents, with no participants reporting memory issues---highlights the need for \amh{future studies with a larger, more diverse population, especially individuals with MCI}. 
\amh{Another limitation of our study is the use of mock AVSs for privacy reasons. Since AVSs vary across healthcare providers—in both format and content—future work should explore the use of image-to-text models to extract relevant information. Additionally, the LLM-powered VA should be designed to handle AVSs in diverse and potentially unstructured formats.}
\amh{Lastly, the potential harms of LLMs---such as hallucinations, bias, and reproducibility issues---in the healthcare context were not explored in our work. Further research is needed to understand and mitigate potential negative impacts of LLMs on users to ensure safe and reliable interactions. Further research is needed to design effective methods to communicate the source of information provided by the LLM-driven personal health assistants.}

\balance

\begin{acks} 
This work was supported by the National Science Foundation award \#1840088 and Malone Center for Engineering in Healthcare. 
\end{acks}

\section*{CRediT author Statement}
\textbf{Amama Mahmood}: Conceptualization, Methodology, Software, Validation, Formal analysis, Investigation, Data curation, Writing - Original draft, Writing - Review \& editing, Visualization, Project Administration. 
\\
\textbf{Shiye Cao}: Conceptualization, Methodology, Validation, Formal analysis, Investigation, Data curation, Writing - Review \& editing 
\\
\textbf{Maia Stiber}: Conceptualization, Methodology, Validation, Formal analysis, Investigation, Data curation, Writing - Review \& editing 
\\
\textbf{Victor Nikhil Antony}: Conceptualization, Methodology, Validation, Formal analysis, Investigation,  Writing - Review \& editing 
\\
\textbf{Chien-Ming Huang}: Conceptualization, Methodology, Resources, Writing - Original draft, Writing - Review \& editing, Visualization, Supervision, Funding acquisition. 

\section*{Declaration of generative AI and AI-assisted technologies in the writing process}

During the preparation of this work the authors used ChatGPT in order to cut down repetitions and improve readability and language. After using this tool/service, the authors reviewed and edited the content as needed and takes full responsibility for the content of the publication.

\bibliographystyle{ACM-Reference-Format}
\bibliography{references}

\newpage

\section*{Appendix}

\begin{table}[hptb!]
\centering
\begin{minipage}{\textwidth}

\caption{Demographics of participants: All participants were fluent in English. All were retired except P10a and W3-B (employed full-time), P11a (employed part-time), and 1C and 2C, who are volunteering or seeking opportunities. Note: ``eq'' means equivalent diploma. P1--P17 participated in the initial interview. W$X$-$I$ represents the co-design workshop participants, where $X$ is the number of the co-design workshop and $I$ is the initial of their preferred name \eg $W1-B$. $XI$ represents evaluation study participants, where $X$ is the participant number and $I$ is the initial of their preferred name \eg $2C$.}
\label{tab:participants}
\centering
\begin{tabular}{lllllllll}
\textbf{P} &\textbf{Gender} & \textbf{Age} & \textbf{Ethnicity} & \textbf{Highest degree} & \textbf{Profession} & \textbf{Disabilities} 
\\
\midrule[1.5pt]
\multicolumn{7}{l}{\textbf{Community Center - Assisted Living (By Self)}} \\
\hline
1   & F & 94   & Caucasian & High school or eq & Homemaker & Memory issues  
\\

2   & M & 66   & Prefer not to say & Bachelor's degree & Physician assistant & Wheelchair  
\\

3   & F & 82   & Caucasian & High school or eq & Accountant clerk & Hearing aid 
\\

4   & M & 79   & Caucasian & Master's degree & Actuary & Wheelchair 
\\
\hline
\multicolumn{7}{l}{\textbf{Community Center - Independent Living (By Self)}} \\
\hline
5   & F & 77   & Caucasian & Bachelor's degree & Social work &  
\\

6; W2-M; 4M  & F & 81   & Caucasian & Master's degree & CIA & Cane/Rollator 
\\

7   & F & 74   & African American & Bachelor's degree & Administrator &  
\\

14  & F & 73   & African American & Bachelor's degree & Senior Claims & 
\\

15; W2-L & F & 73   & African American & Master's degree & Educator/Pastor &  
\\
W2-E & F & 79 & Caucasian &  Bachelor's degree & Admin Assistant & Walker
\\
\hline
\multicolumn{7}{l}{\textbf{Community Center - Independent Living (With Spouse)}} \\
\hline
W3-N & F & 81 & Asian & Trade school & Teacher & \\
W3-Y & M & 88 & Asian & Master's degree & Teacher & \\
\cdashline{1-7}
1C & F & 84 & Caucasian & Trade school & Tailor & \\
\hline
\multicolumn{7}{l}{\textbf{Homeowner (By Self)}} \\
\hline
8   & F & 84   & Caucasian & Bachelor's degree & Writer/editor & Walker 
\\

9; W1-B   & M & 75   & Caucasian & Bachelor's degree & Case management & 
\\

13  & M & 81   & Caucasian &  &  &  Vision issues 
\\

17  & F & 68   & Caucasian & Trade school & EEG technologist & 
\\
W1-P  & F & 81   & Caucasian & Master's degree & Teacher/social worker & 
\\
W1-L; 4L  & F & 81   & Caucasian & Bachelor's degree & Dietitian & Mild arthritis
\\
2C & F & 76 & Caucasian & PhD & Neuroscientologist & Vision issues\\

\hline
\multicolumn{7}{l}{\textbf{Homeowner (Couple, With Spouse)}} \\
\hline

10a & M & 68   & Caucasian & Bachelor's degree & Staff engineer &  
\\

10b & F & 94   & Caucasian & Master's degree & Education & Walker/scooter 
\\
\cdashline{1-7}
11a; W1-R; 5R & M & 72   & Caucasian & Master's degree & Urban planning & 
\\

11b & F & 71   & Asian & Master's degree & Social work/teacher & 
\\
\cdashline{1-7}
12a & M & 75   & Caucasian & Bachelor's degree &  &  
\\

12b & F & 72   & Caucasian & Nursing diploma & Registered nurse &  
\\
\cdashline{1-7}
16a & M & 76   & Caucasian & Bachelor's degree & Business owner &  
\\

16b & F & 75   & Caucasian & Nursing diploma & Nurse & 
\\
\cdashline{1-7}
W3-B & F & 65 & Asian & Ph.D. & Lecturer & 
\\

\hline
\end{tabular}
\end{minipage}

\end{table}

\begin{table*}[!h]
\centering
\caption{Participants for co-design workshops: Voice assistant technology use and current health management practices. For medication intake they rely on memory unless states otherwise.} 
\label{tab:tech-use}
\begin{tabular}{l l l l}

\textbf{P} & \textbf{VA technology use}& \textbf{Appointments} & \textbf{Medications} \\

\midrule
\midrule
\multicolumn{4}{l}{\textbf{Workshop 1: Community dwelling adults}}\\
\hline
W1-B   & Uses smart speaker based VA (Alexa) & Phone calendar & Relies on memory\\
W1-P   & Uses smart speaker based VA (Alexa) & Paper calendar & Pill boxes, one for morning and one for night \\
W1-R & Used smart speaker based VA (Alexa) & Phone calendar & Relies on memory\\
W1-L   & Uses phone based VA (Siri) &  Paper calendar & Pill boxes, one for morning and one for night\\
\hline
\multicolumn{4}{l}{\textbf{Workshop 2: Independent living community center}}\\
\hline
W2-L& Uses multiple smart speaker based VA (Alexa) & Phone calendar, VA reminders & Pill boxes, sometimes use Alexa for reminders\\
W2-M & Used smart speaker based VA (Alexa) & Paper calendar & Pill boxes\\
W2-E   & Never used VA &  Phone calendar, no reminders & Pill boxes\\
\hline
\multicolumn{4}{l}{\textbf{Workshop 3: Immigrants }}\\
\hline
W2-B & Uses smart speaker (Apple's HomePod) & Phone calendar, reminders & Relies on memory \\
W2-N & Never used VA & Paper calendar & Pill boxes, one for morning and one for night \\
W2-Y & Never used VA & Paper calendar & Pill boxes, one for morning and one for night \\
\hline
\end{tabular}
\end{table*}

\begin{table*}[!h]
\centering
\caption{Example of profile creation for participant 1C from the conversation.}
\label{tab:conversations-persona}
\begin{tabular}{X S}
\textbf{VA} & \textbf{Participant} \\
\midrule
\midrule
Now, is your daily routine like wake-up time, meals, and sleep time consistent, or does it change? Also, do you have the same routine over the weekend? \circlewithnumber{custompurple}{5}{0.8} \circlewithnumber{custompurple}{6}{0.8}& Yes, that's my routine you mentioned and I do keep the same routine over the weekend. \\
\cdashline{1-2}
Got it. How do you store and manage your medications? Do you use pill boxes or keep them in their original containers? \circlewithnumber{custompink}{2}{0.8} \circlewithnumber{custompurple}{4}{0.8}  \circlewithnumber{custompurple}{6}{0.8} & I keep a week supply in a pill box dated and I take them at one time in the evening just before I go to bed and the rest the bottles are in my nightstand my night table and I fill up the pill box weekly and that's about all I can say and that subject. \\
 \cdashline{1-2}
Thanks for sharing that. To set the reminders correctly, could you tell me what time you usually wake up and go to bed? \circlewithnumber{custompink}{7}{0.8} \circlewithnumber{custompink}{8}{0.8} & I go to bed between nine and ten and I wake up between six and seven. \\
 \cdashline{1-2}
Thanks for that. When do you usually have your meals, like breakfast, lunch, and dinner? \circlewithnumber{custompink}{8}{0.8} & Breakfast when I'm hungry lunch around one. \\
 \cdashline{1-2}
Thanks for sharing. Could you also let me know when you usually have dinner? \circlewithnumber{custompink}{8}{0.8} & Between five and six p.m.\\
 \cdashline{1-2}
I missed your response. Could you please repeat when you usually have dinner? & Dinner is usually between five and six p. m.  \\
 \cdashline{1-2}
Got it. Is there any other specific preference you have regarding medication reminders? & I need the pill box that's all thank you \\
\multicolumn{2}{p{17.5cm}}{ \textbf{VA generated user profile:} user has consistent routine, wakes up between 6 and 7 AM, and goes to bed between 9 and 10 PM. Breakfast is variable, lunch is around 1 PM, and dinner is between 5 and 6 PM. Manages medications using a weekly pill box, taken all at one time in the evening just before going to bed. Keeps the rest of the bottles in a nightstand.}\\
\hline

\end{tabular}
\end{table*}


\end{document}